\documentclass{article}
\usepackage{amsmath,amssymb}
\usepackage{epsfig,multirow}
\usepackage{bigstrut}
\usepackage{array}
\usepackage{float}
\usepackage[a4paper,top=3cm,bottom=2cm,left=3cm,right=3cm,marginparwidth=1.75cm]{geometry}
\usepackage{lineno}
\usepackage{jcappub} 
\usepackage{academicons}

\title{Impact of the Magnetic Horizon on the Interpretation of the Pierre Auger Observatory Spectrum and Composition Data}

\author[13]{A.~Abdul Halim,}
\author[73]{P.~Abreu,}
\author[55,53]{M.~Aglietta,}
\author[1]{I.~Allekotte,}
\author[81,80,71]{K.~Almeida Cheminant,}
\author[7,12]{A.~Almela,}
\author[46,47]{R.~Aloisio,}
\author[79]{J.~Alvarez-Mu\~niz,}
\author[79]{J.~Ammerman Yebra,}
\author[59,48]{G.A.~Anastasi,}
% ...

\affiliation[1]{Centro At\'omico Bariloche and Instituto Balseiro (CNEA-UNCuyo-CONICET), San Carlos de Bariloche, Argentina}
% ...
\affiliation[7]{Instituto de Tecnolog\'\i{}as en Detecci\'on y Astropart\'\i{}culas (CNEA, CONICET, UNSAM), Buenos Aires, Argentina}
% ...
\affiliation[12]{Universidad Tecnol\'ogica Nacional -- Facultad Regional Buenos Aires, Buenos Aires, Argentina}
\affiliation[13]{University of Adelaide, Adelaide, S.A., Australia}
% ...
\affiliation[46]{Gran Sasso Science Institute, L'Aquila, Italy}
\affiliation[47]{INFN Laboratori Nazionali del Gran Sasso, Assergi (L'Aquila), Italy}
\affiliation[48]{INFN, Sezione di Catania, Catania, Italy}
% ...
\affiliation[53]{INFN, Sezione di Torino, Torino, Italy}
% ...
\affiliation[55]{Osservatorio Astrofisico di Torino (INAF), Torino, Italy}
% ...
\affiliation[59]{Universit\`a di Catania, Dipartimento di Fisica e Astronomia ``Ettore Majorana``, Catania, Italy}
% ...
\affiliation[71]{Institute of Nuclear Physics PAN, Krakow, Poland}
% ...
\affiliation[73]{Laborat\'orio de Instrumenta\c{c}\~ao e F\'\i{}sica Experimental de Part\'\i{}culas -- LIP and Instituto Superior T\'ecnico -- IST, Universidade de Lisboa -- UL, Lisboa, Portugal}
% ...
\affiliation[79]{Instituto Galego de F\'\i{}sica de Altas Enerx\'\i{}as (IGFAE), Universidade de Santiago de Compostela, Santiago de Compostela, Spain}
\affiliation[80]{IMAPP, Radboud University Nijmegen, Nijmegen, The Netherlands}
\affiliation[81]{Nationaal Instituut voor Kernfysica en Hoge Energie Fysica (NIKHEF), Science Park, Amsterdam, The Netherlands}
% ...
\affiliation[c]{also at University of Bucharest, Physics Department, Bucharest, Romania}
% ...

\emailAdd{auger\_spokespersons@fnal.gov}

\abstract{
The flux of ultra-high energy cosmic rays reaching Earth above the ankle energy (5 EeV) can be described as a mixture of nuclei  injected by extragalactic sources with very hard spectra and a low rigidity cutoff.
Extragalactic magnetic fields existing between the Earth and the closest sources can affect the observed CR spectrum by reducing the flux of low-rigidity particles reaching Earth. We perform a combined fit of the spectrum and distributions of depth of shower maximum measured with the Pierre Auger Observatory including the effect of this magnetic horizon in the propagation of UHECRs in the intergalactic space.
We find that, within a specific range of the various experimental and phenomenological systematics, the magnetic horizon effect can be relevant for turbulent magnetic field strengths in the local neighbourhood in which the closest sources lie
of order $B_{\rm rms}\simeq (50-100)\,{\rm nG}\,(20\,\rm{Mpc}/{d_{\rm s})( 100\,\rm{kpc}/L_{\rm coh}})^{1/2}$, with $d_{\rm s}$ the typical intersource separation and $L_{\rm coh}$ the magnetic field coherence length. When this is the case, 
the inferred slope of the source spectrum becomes softer and can be closer to the expectations of diffusive shock acceleration, i.e., $\propto E^{-2}$.
An additional cosmic-ray population with higher source density and softer spectra, presumably also extragalactic and dominating the cosmic-ray flux at EeV energies, is also required to reproduce the overall spectrum and composition results for all energies down to 0.6~EeV.}

\begin{document}
\maketitle

\flushbottom

\section{Introduction}

To understand the properties of the sources of cosmic rays (CRs) required in order to account for the spectrum and composition inferred from data collected at the Pierre Auger Observatory, combined fits to the measurements of the CR flux and distribution of the depth of shower maximum ($X_{\rm max}$) were performed  considering different astrophysical source scenarios \cite{aa17,xcf} (see also \cite{Aloisio:2013hya,gl15,AlvesBatista:2018zui,Heinze:2019jou}). These fits adopted  continuous distributions of CR sources, eventually allowing for a redshift evolution of their emissivities. The emitted particles were then propagated including the attenuation effects due to the CR interactions with the background radiation, and the resulting fluxes at the Earth were compared to observations in order to obtain the best fitting source parameters. One source population dominates the fluxes above the ankle energy ($\sim 5$~EeV), while a second source population is required to also explain the observations below the ankle, as in particular shown in \cite{Aloisio:2013hya,gl15,xcf}, with each population having sources with specific spectral and composition properties. Since the observations indicate that the CR composition becomes increasingly heavier above the ankle energy \cite{ayus19,comp2014}, this can naturally result if the different mass components have cutoffs which depend on the  rigidity~$R=E/Z$, i.e.\ cutoff energy proportional to the atomic number $Z$, as  expected from electromagnetic acceleration processes\footnote{We will refer to the quantity $E/Z$ as rigidity, measured in eV, while the actual magnetic rigidity is $pc/(eZ)\simeq E/(eZ)$ and is measured in Volts.}, so that the light components do not reach the highest energies. Moreover, in order for the flux of the heavier components dominating at the highest energies to be sufficiently suppressed below the ankle, where a light composition is inferred, the individual spectral shapes of the different elements contributing to the high-energy population need to be quite hard. In particular, considering that below the cutoffs the source spectra have a power-law shape $E^{-\gamma}$,  values of $\gamma<1$ are inferred (and in some cases as low as $-2$) for the high-energy population, which is in tension with the value $\gamma \simeq  2$ that is expected from diffusive shock acceleration (DSA). Similar values for the spectral index of the high-energy population were obtained when fitting the spectrum, composition and arrival direction distribution at energies above $10^{19}$\,eV \cite{adcfit}. The energy dependent magnetic confinement of heavy nuclei and their photodisintegration in the source environment has been considered as a possible explanation of the suppression of the flux of the high-energy component at low energies in \cite{Globus:2014fka,gl15,un15,bi18,Supanitsky:2018jje,Muzio:2019leu,he20,co23}, leading to an emission spectrum harder than that produced in the acceleration process.
The population dominating the flux below the ankle is instead required to have a very soft spectrum ($\gamma>3$) and a mix of protons and intermediate-mass nuclei. The very soft spectrum may eventually be  due to the superposition of many  sources with a harder spectrum and a distribution of cutoff rigidities \cite{ka06}. 
The proton component below the ankle could alternatively result from interactions in the source environment of the intermediate and heavy-mass nuclei from the high-energy population, producing nucleons, among which neutrons that can escape the magnetized medium and decay into protons in flight. This model has been tested already in \cite{xcf,luce22}.

The previous combined fits of the spectrum and composition
\cite{aa17,xcf} have considered a continuous distribution of CRs sources in the universe. In the more realistic case of a discrete source distribution, the presence of sizeable extragalactic magnetic fields between the closest sources and the Earth can affect the shape of the spectrum of cosmic rays reaching us. In particular, an alternative explanation for the hardness of the high-energy population spectrum is that it is due to a magnetic horizon effect which, as a consequence of the CR diffusive propagation through the intergalactic turbulent magnetic fields, could suppress the flux at low energies \cite{al04,al11,mo13}. This suppression  is relevant when the time for CR diffusion from the closest sources becomes longer than the age of the sources, so that the low-energy cosmic rays have not enough time to reach the Earth, and for this to happen one then needs  strong magnetic fields as well as relatively large intersource distances (i.e. small source densities). Following the ideas in \cite{al04,mo13,wit17,mo20,go23}, we here extend the combined fit analysis of the spectrum and composition data by including the magnetic horizon effects associated to the finite intersource separations and the presence of extragalactic magnetic fields, which are two ingredients that should be included in a more realistic description of reality. We explore in particular whether the resulting suppression of the flux at low energies can  allow for softer source spectra so as to alleviate  the existing tension with the expectations from DSA. We don't consider the Galactic magnetic field since it is not expected to significantly modify the spectrum of extragalactic cosmic rays. 

\section{The source populations}

The general framework we consider in this study to reproduce the energy spectrum both below and above the ankle feature is that already explored in  \cite{xcf}, consisting of two different source populations. One source population dominates the CR flux above a few EeV while a second one dominates at lower energies.
For each of these populations the source spectra  of the different mass components are considered to be  described as power laws having a rigidity dependent cutoff which strongly suppresses the fluxes above an energy $ZR_{\rm cut}$. 
 This kind of cutoffs are expected when the acceleration has an electromagnetic origin, so that the maximum energy achieved is proportional to the charge of the particle, and when there are no strong  interactions in the acceleration region.\footnote{ A different situation could be that in which the maximum energy would result from interaction effects, since for instance nuclei could be limited by photodisintegration processes which do not affect  the protons. On the other hand, scenarios in which the sources have a distribution of cutoff rigidities have been discussed in \cite{foteini}.}
In that case, the differential particle generation rate of each component of atomic number $Z$ and mass number $A$, per unit volume, energy and time, is
\begin{equation}
    \tilde Q_{A}^a(E,z)=\tilde Q_0^a\xi^a(z) f^a_A\left(\frac{E}{E_0}\right)^{-\gamma_a}F_{\rm cut}\left(\frac{E}{ZR^a_{\rm cut}}\right),
    \label{qa.eq}
\end{equation}
with $a = ({\rm L, H})$ identifying the population dominating at low and high energies respectively. For each population, the normalization $\tilde Q^a_0$ is the present total differential rate of CR emission per unit energy, volume and time, at the reference energy $E_0$ (smaller than the hydrogen cutoff  $R^a_{\rm cut}$ and taken here as 1\,EeV), at which the relative source fractions of the different elements are $f^a_A$. We consider that five representative elements are  emitted at the sources: H, He, N, Si and Fe. 

The factor $\xi^a(z)$ parameterises the evolution of the emissivity as a function of the redshift $z$, for which we consider here two possibilities. The first is that of non-evolving sources  (NE) having   $\xi^a_{\rm NE}(z)=1$, and the other assumes that the emissivities scale with the star formation rate (SFR) as parameterised in \cite{ho06}, i.e. 
\begin{equation}
        \xi^a_{\rm SFR}(z) = \begin{cases}
        (1+z)^{3.44}, \quad &\text{if} \ z \leq 0.97; \\
        12.3\, (1+z)^{-0.26}, \quad &\text{if} \ 0.97 \leq z\leq 4.44, \\
        \end{cases}
    \end{equation}
    with a steep decline at higher redshifts. 
We will actually consider maximum redshifts $z_{\rm max}=1$ for the NE case and $z_{\rm max}=4$ for the SFR case, since the contribution from CR sources farther away is negligible. For the cutoff function $F_{\rm cut}$ we adopt the following parametrization  \cite{xcf}
\begin{eqnarray}
  F_{\rm cut}(y)= {\rm sech}(y^{\Delta}), \nonumber
\end{eqnarray}
 with the parameter $\Delta$ determining the steepness of the cutoff shape, which turns out to have a significant impact on the fit. We will consider the three representative values $\Delta = 1, 2 $ and 3.\footnote{The case $\Delta=1$ is very close to the broken exponential cutoff case considered in \cite{xcf}. }

The particles emitted at the sources are  propagated up to the Earth using the SimProp~v2r4 software \cite{al17}. The resulting fluxes depend on the nuclear photo-disintegration cross sections, as well as on the extragalactic background light  model considered to evaluate the interactions during propagation. We adopt  in the analysis the  photodisintegration cross section from TALYS \cite{talys} and the extragalactic background radiation from Gilmore et al. \cite{gi12}.  
The inferred composition  depends on  the hadronic model used to interpret the $X_{\rm max}$ measurements, which was found in  \cite{aa17,xcf} to  be the
assumption that mostly affected the values of the fitted fractions. 
We will explore the dependence of the results on this by considering both EPOS-LHC \cite{epos} and Sibyll~2.3d \cite{sibyll}   hadronic interaction  models.

\section{The magnetic horizon effect}

An alternative explanation \cite{mo13} for the hardness of the observed high-energy component spectra is that it is not due to the hardness of the source spectra, but to a magnetic horizon effect \cite{lemoine,be07,bere08} which, as a consequence of the CR diffusive propagation through the intergalactic turbulent magnetic fields, could increasingly suppress their flux for decreasing energies.  We will here consider the case of steady sources,  and given the large $z_{\rm max}$ values considered  the characteristic time for the source emission is of the order of the age of the universe. Note that the suppression  effects at low energies could be more pronounced if the sources were transient rather than steady, in which case smaller strength of the magnetic fields and/or smaller intersource distances would be required, but in this case the results will depend on the specific emission histories of the different nearby CR sources and their distances \cite{mo19,ei23}.

Magnetic fields are known to permeate the Universe, having different strength on different scales, and they hence affect the propagation of the charged cosmic rays.  In our  galaxy they have a strength of several \textmu{G} and extend over scales up to tens of kpc. They can be modelled using several components, such as the regular ones in the disk and in the halo which are coherent over kpc scales, or the random and striated turbulent components which have a typical coherence length of about 50~pc (see e.g. \cite{haverkorn15}). These components will deflect the incoming UHECR and also lead to a diffusive behaviour for the CRs with energies below about 0.1$Z$~EeV.  In clusters of galaxies the turbulent magnetic fields can reach values of 1 to 10~\textmu{G}, with typical coherence lengths of 1 to 100~kpc \cite{fe12,va11}. In large-scale structure filaments the magnetic fields are more uncertain, but may well have strengths in the 10 to few 100~nG range, with coherence lengths in the 10~kpc to 1~Mpc range \cite{xu06,va11,va17}, while in the voids of the large-scale structures they are expected to be much weaker, with strengths smaller than 1~nG  \cite{du13,pshirkov16}.  The strength of these fields depends on the mechanism producing them. In particular, they could result  from the amplification via flux conserving gravitational compression of primordial seeds,  such as those left over from an inflationary period or those produced in phase transitions in the early universe, or alternatively they could result from outflows of galactic fields that were amplified by some dynamo process.

The presence of extragalactic magnetic fields  (EGMF)
can be the dominant contributor  to a magnetic horizon effect and may hence affect the observed cosmic-ray spectrum. In particular, EGMF are expected to be enhanced in our local neighbourhood  where the closest UHECR sources should lie, such as within the Local Supercluster which has a typical radial extent of about 20~Mpc. Also note that the assumption of  equipartition between the energy in thermal gas motion and in the magnetic field   leads to an estimation of the magnetic field strength within filaments and sheets of the large scale structure of order $10^2$\,nG \cite{ry98}.
We will model for simplicity the EGMF as being turbulent and isotropic, parameterized by its root-mean-square  amplitude ($B_{\rm rms}$) and  coherence length ($L_{\rm coh}$), considering a Kolmogorov spectrum for the turbulence. A critical energy can be defined as that for which the effective Larmor radius associated to $B_{\rm rms}$ equals the coherence length, and for a particle of atomic number $Z$ it is given by  $E_{\rm crit} \equiv |e|ZB_{\rm rms}L_{\rm coh}\equiv ZR_{\rm crit}$, with the critical rigidity  
\begin{equation}
R_{\rm crit} \equiv  |e|B_{\rm rms}L_{\rm coh} \simeq 0.9\ (B_{\rm rms}/{\rm nG})(L_{\rm coh}/{\rm Mpc}) \,{\rm EeV},
    \label{Rcrit}
\end{equation}
being the critical energy for a H nucleus. The critical energy separates two different propagation regimes. The first one is that  of resonant diffusion, at energies $E\ll E_{\rm crit}$, in which deflections are large even before traversing a distance  $L_{\rm coh}$. The second one is for energies $E\gg E_{\rm crit}$, in which case the propagation is instead  quasirectilinear within a coherence length and the diffusive regime is only attained after traversing much longer distances. 

We will  consider a distribution of UHECR sources with uniform density $n_{\rm s}$, with the average distance between them being $d_{\rm s}= n_{\rm s}^{-1/3}$.  
When the particle rigidities are low enough so that  the  CR diffusive travel time from the closest sources in the presence of an EGMF becomes larger than the source age,  their flux will get significantly suppressed. 
It is useful to quantify this effect through a suppression function  $G(E) \equiv J(E)/J(E)_{d_{\rm s} \rightarrow 0}$, given by the ratio between the actual flux at Earth from the discrete source distribution to the flux that would result in the limit of a continuous source distribution (with the same emissivity per unit volume).\footnote{For the case of a continuous source distribution, the magnetic fields  have no suppression effect due to the so-called {\em propagation theorem} \cite{al04}, which reflects the fact that the suppression of the faraway sources gets compensated by the diffusive enhancement of the nearby ones.  }
Using simulations of the  propagation of particles in turbulent magnetic fields with a Kolmogorov spectrum performed in an extended implementation of the SimProp code where both interactions with radiation fields and magnetic deflections are accounted for, the suppression function has been parameterised as \cite{go21}
\begin{equation}
    G(x) = \exp \left[- \left( \frac{a \,X_{\rm s}}{x + b\,(x/a)^\beta}\right)^{\alpha} \right],
    \label{gfactor}
\end{equation}
where  $x\equiv E/(ZR_{\rm crit})$ and $X_{\rm s} = d_{\rm s}/\sqrt{r_HL_{\rm coh}}$ is the normalized intersource distance, with $r_H = c/H_0$ the Hubble radius (in terms of the speed of light $c$ and the present day Hubble constant $H_0\simeq 70$\,km\,s$^{-1}$\,Mpc$^{-1}$), so that 
\begin{equation}
    X_{\rm s} \simeq \frac{d_{\rm s}}{10\,{\rm Mpc}}\sqrt{\frac{25\,{\rm kpc}}{L_{\rm coh}}}.
    \label{eq.xs}
\end{equation}
The parameters  $a,\, b,\, \alpha$ and $\beta$  are sensitive to the distribution of the initial redshifts of the particles that reach the Earth. They hence depend on  the assumed cosmological evolution of the source population emissivity  as well as on whether the particles are primaries emitted at the sources or secondaries produced by photo-disintegration interactions during their propagation.  The suppression also depends slightly on the spectral index of the sources, and the values of the parameters obtained in ~\cite{go21} are tabulated in the Appendix~A, both for the NE and SFR scenarios. 

The spectra of the different mass components reaching the Earth in the presence of EGMF can be obtained as the product of those in the absence of magnetic fields times the corresponding suppression factor $G$. Thus, the magnetic horizon effect is  accounted through two parameters: the critical rigidity  $R_{\rm crit}$ and the normalized intersource distance $X_{\rm s}$. This description allows us to probe a wide range of values of the magnetic field amplitudes and coherent lengths as well as of the source densities.  
It is important to keep in mind that due to the diffusion the contribution from the faraway sources becomes strongly suppressed for decreasing rigidities and eventually the flux at low energies is dominated by that from the nearby sources. Thus, the magnetic field which is relevant for the suppression due to the magnetic horizon  effect is the one between the closest sources and the observer, with the closest source being on average at a distance of 0.55$d_{\rm s}$ \cite{mo13}.\footnote{Note that Eq.\,(\ref{gfactor}) assumes an uniform magnetic field distribution in all space. If we were to consider instead that the magnetic field is contained within a finite spherical region around the observer, the suppression will still be well described by Eq.\,(\ref{gfactor}) as long as  
the closest source lies inside this region, i.e. for $d_{\rm s}$ not larger than the bubble diameter. The flux would however become less suppressed  if the closest source lies outside  the region containing the magnetic field. }

If one considers magnetic field amplitudes in the range $1\,{\rm nG}\lesssim B_{\rm rms}\lesssim 200$\,nG and coherence length such that $25\,{\rm kpc}\lesssim L_{\rm coh}\lesssim 1$\,Mpc, one expects that $0.022\,{\rm EeV}\lesssim R_{\rm crit}\lesssim 180$\,EeV. Moreover,  if the intersource average distance is in the range $4\,{\rm Mpc}\lesssim d_{\rm s}\lesssim 40$\,Mpc  one should have that $0.05\lesssim X_{\rm s}\lesssim 4$. We will hence consider parameters  $R_{\rm crit}$ and $X_{\rm s}$  within these ranges when performing the fits that include the magnetic horizon effects. 

\begin{figure}[t]
    \centering
    \includegraphics[width=0.49\textwidth]{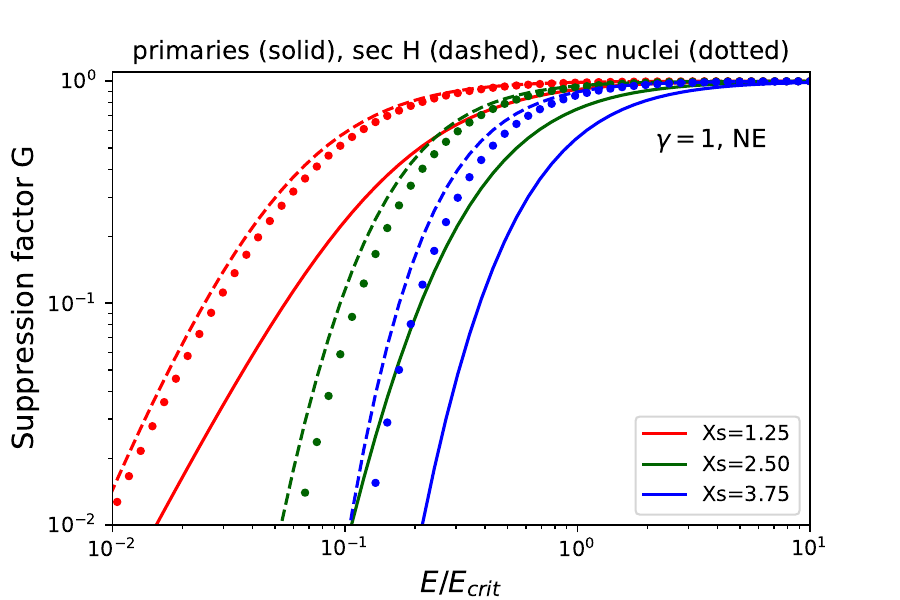}
    \includegraphics[width=0.49\textwidth]{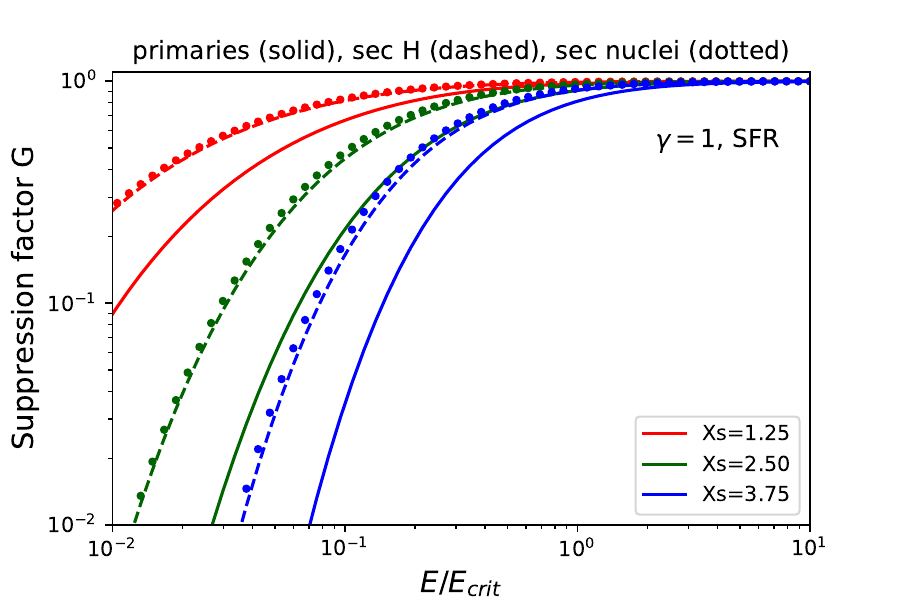}
    \caption{Magnetic suppression factor as a function of energy over critical energy for three values of the normalized intersource distance $X_{\rm s}$. The left panel assumes a no evolution scenario while the right panel assumes a star formation rate  one. Results are shown for primaries (solid lines), secondary protons (dashed
lines) and intermediate secondary nuclei (dotted lines).}
    \label{fig:Gfactor}
\end{figure}

The magnetic suppression factor $G$ is displayed in Fig.~\ref{fig:Gfactor}  as a function of $E/E_{\rm crit}$, for different values of $X_{\rm s}$ and for the NE  (left) and SFR (right) evolutions, considering $\gamma=1$. 
The implied strong hardening of the spectrum at low energies  is apparent, with the suppression occurring in general  in the regime of resonant diffusion, i.e. for energies below  $ E_{\rm crit}\equiv Z R_{\rm crit}$. 
This effect is clearly rigidity dependent, given its magnetic origin.  One also finds that for increasing $X_{\rm  s}$ the suppression appears for larger values of $E/E_{\rm crit}$. Hence,  increasing $X_{\rm s}$ and simultaneously  decreasing  $E_{\rm crit}$ the different curves can lead to comparable suppressions as a function of the energy, what will lead to some approximate degeneracies between these parameters,  although the shape of the suppression becomes steeper  for increasing values of $X_{\rm s}$. For the same values of the parameters $X_{\rm s}$ and $E_{\rm crit}$ the suppressions are milder for the SFR case than for the NE one, given that due to the enhanced emission at high redshifts  the particles have on average more time to arrive from their sources in the first case (note that for diffusing particles the redshift is a measure of the time travelled rather than of the source distance).
We show the results both for primaries (solid lines), secondary protons (dashed lines) and intermediate secondary nuclei (dotted lines), where the primaries are those in which the detected nucleus is in the same mass group as the emitted one, considering the different mass groups as those with values of $A$ of 1 (H), 2--4 (He), 5--16 (N), 17--30 (Si) and 31-56 (Fe). The intermediate secondary nuclei are those in a lighter mass group than the primary injected particle. The milder suppression of the secondaries is also understood because their average redshift of production in photodisintegration processes is higher than the average production redshift of the primaries, given that the photodisintegrations processes get enhanced at higher redshifts. Also note that the suppression of the secondary protons is quite similar to the suppression of the secondary nuclei of the same rigidity.

In the scenarios with two source populations that we are considering,  we will assume that the low-energy component providing the bulk of the flux at few EeV energies arises from sources which are  more abundant than those of the high-energy  component,  having then a smaller intersource distance (smaller $X_{\rm s}$). The magnetic suppression will in this case  affect the LE component  at energies much lower than the HE component \cite{lemoine,blasi15}, and we will hence neglect the magnetic horizon effects on the low-energy component.  Since this approximation may become less accurate for decreasing energies and, in addition, a non-negligible Galactic contribution of heavy elements may also extend above the second-knee feature present at 0.1~EeV, the model predictions at energies below  1\,EeV are expected to be affected by those effects. 

\section{Combined fit to the spectrum and composition}

\subsection{Flux model }

The flux of particles  reaching the Earth from each population of sources $a$ can then be obtained for a given model of the sources and extragalactic magnetic field parameters as
\begin{equation}
    J^{a}_{\rm mod}(E')=\frac{c}{4\pi}G\left(\frac{E'}{E_{\rm crit}}\right)\sum_{A,A'}\int_0^{z_{\rm max}}{\rm d}z\, \left|\frac{{\rm d}t}{{\rm d}z}\right| \int{\rm d}E\,  \tilde Q^{a}_A(E,z)\frac{{\rm d}\eta_{A',A}(E',E,z)}{{\rm d}E'},
    \label{fluxearth}
\end{equation}
with 
\begin{equation}
    \left|\frac{{\rm d}t}{{\rm d}z}\right|=\frac{1}{(1+z)H_0\sqrt{\Omega_\Lambda+(1+z)^3\Omega_m}},
\end{equation}
where  $E'$ and $E$ denote the energy observed at Earth and that at the sources, respectively,   $\Omega_\Lambda\simeq 0.7$ and $\Omega_m\simeq 0.3$. 
 The effect of the interactions with background photons is accounted for by ${\rm d} \eta_{A',A}(E',E, z)/{\rm d} E'$, which represents the differential probability that particle with energy $E'$ and mass number $A'$ reaches $z =0$ when a particle  with energy $E$ and mass $A$ is injected at a redshift $z$, and this quantity  is obtained from simulations computed using SimProp. 

\subsection{Fit procedure} 
\label{fit}

The  parameters of the model are obtained by maximising the likelihood so that the assumed model reproduces the observed data. To do so, we follow the steps outlined in \cite{aa17,xcf}. The test statistic used to parameterise the goodness of fit is the deviance, which is defined as twice the negative logarithm of the likelihood ratio between each model and one that would perfectly describe the observed data (called the \textit{saturated model}), $D = -2 \ln \left({L}{/L_{\rm sat}}\right)$. This implies that for the parameters at which the deviance is at a minimum the likelihood of the model is maximized. 

Since both the energy spectrum and $X_{\rm max}$ distributions were measured independently, the complete likelihood of our model is simply the product $L_J \cdot L_{X_{\rm max}}$. Here $L_J$ represents the likelihood of the energy spectrum, defined as
\begin{equation}
 L_J = \prod_i \frac{1}{\sqrt{2\pi }\sigma_i} \exp \left[  -\frac{( J_i^{\rm obs} - J_i^{\rm mod} )^2}{2\sigma_i^2} \right], 
\end{equation}
where $i$ denotes the i-th detection energy bin; $J_i^{\rm obs}$ and $J_i^{\rm mod}$ are the observed flux and the one predicted by the model, respectively, and $\sigma_i$ the associated uncertainties in the measurements. Meanwhile, $L_{X_{\rm max}}$ represents the likelihood of the $X_{\rm max}$ distribution measurements, following the expression

\begin{equation}
   L_{X_{\rm max}} = \prod_i n_i^{\rm obs}! \prod_j \frac{(G^{\rm mod}_{i,j})^{ k^{\rm obs}_{i,j} }}{k^{\rm obs}_{i,j}!\,}
\end{equation}
where $k^{\rm obs}_{i,j}$ is the number of events in the i-th energy bin and the j-th bin of the  $X_{\rm max}$ distribution, $n_i^{\rm obs}$ is the total number of event in the i-th energy bin, and $G^{\rm mod}_{i,j}$ are the corresponding model predictions obtained from the modified Gumbel functions which  account for the detection and resolution effects as described in \cite{xcf}.

The deviance is minimized using the Minuit library \cite{Minuit}. The uncertainties in the spectral parameters, fractions and magnetic horizon parameters are obtained via the MINOS procedure included in the Minuit package. These correspond to the change in each parameter for which the deviance increases by one unit when minimising with respect to the rest of the parameters. The uncertainties in the source emissivities are obtained via Monte Carlo simulations in which the different parameters are allowed to vary within their uncertainties.

\subsection{Data sets}

We fit the energy spectrum determined by the Pierre Auger Observatory
using the events detected by the Surface Detector array. The array  with stations separated by 1500~m is used above 2.5~EeV while the denser array with stations separated by 750~m is used for smaller energies \cite{spectrum}. The energy range considered covers from $10^{17.8}$ eV to $10^{20.2}$ eV, in logarithmic bins of width $\Delta \log_{10} (E/{\rm eV})=0.1$. For the composition we fit the $X_{\max}$ distributions measured using the Fluorescence Detector telescopes in the energy range from $10^{17.8}$ eV to $10^{19.6}$ eV  in logarithmic bins of width $\Delta \log_{10} (E/{\rm eV})=0.1$, plus one additional bin including the events with energies above $10^{19.6}$ eV \cite{ayus19}. The  $X_{\max}$ distributions are binned in bins of width $\Delta X_{\rm max}=20$\,g\,cm$^{-2}$ in each energy interval.  These are the same data sets considered in \cite{xcf,go23}.

\subsection{Results}

\subsubsection{Results obtained in the absence of magnetic fields}

As a starting point, we first present in Table \ref{table_0} the results of the fit performed in the absence of magnetic fields and for non-evolving sources,  for the different cutoff functions considered and for both EPOS-LHC and Sibyll~2.3d hadronic interaction models, in line with the analysis in \cite{xcf}, and with compatible results.\footnote{Minor changes in the fitted parameters appear due to the  inclusion of a local overdensity in the analysis of \cite{xcf}.}

The spectrum parameters ($\gamma$ and $R_{\rm cut}$) of the high-energy and low-energy populations are reported, as well as the  obtained deviance.  It is interesting to note the effect of the cutoff function  on the spectral parameters of the HE component. Sharper cutoffs (larger $\Delta$) prefer softer spectra (larger $\gamma$) and a larger rigidity cutoff at the sources, for both hadronic models. Actually, the different parameters in each case combine to  produce a similar injection spectral shape at the sources in the most relevant energy range, as it can be seen in Fig.~\ref{fig:injection_noB} in Appendix~B. The  deviance  obtained is smaller for the milder ($\Delta =1$) cutoff, although requiring a very hard spectrum,  and the deviance grows for increasing $\Delta$, which lead to softer spectral indices. For all the cases considered we obtain that $\gamma_{\rm H} < 1 $, and in particular for $\Delta=1$ one has that $\gamma_{\rm H}<-1.67$. For a given value of $\Delta$ the fits using EPOS-LHC lead  to smaller deviances than those considering Sibyll~2.3d, while these latter lead to larger values of $\gamma_{\rm H}$.
The cutoff of the low-energy population often slides to the maximum allowed value in the fit (100~EeV),  although the deviance is quite degenerate for values of $R_{\rm cut}^{\rm L}$ larger than 25~EeV since the LE population makes a subdominant contribution at those energies. We report in these cases the range of $R_{\rm cut}^{\rm L}$ leading to a deviance within one unit from the value at the boundary.

\begin{table}[t]
\centering
{\small
\begin{tabular}[H]{ @{}c| c c c c c| c c  c c c @{}}
 \multicolumn{11}{c}{no EGMF, NE-NE}
\\
\hline
& \multicolumn{5}{c|}{EPOS-LHC} & \multicolumn{5}{c}{Sibyll~2.3d}  \\
\hline
$\Delta$ & $\gamma_{\rm H}$ & $R_{\rm cut}^{\rm H}$ & $\gamma_{\rm L}$ & $R_{\rm cut}^{\rm L}$ & $D$& $\gamma_{\rm H}$ & $R_{\rm cut}^{\rm H}$ & $\gamma_{\rm L}$ & $R_{\rm cut}^{\rm L}$  & $D$\bigstrut[t]\\ 
& &  [EeV] & &  [EeV] &($N=353$)  & &  [EeV] & &  [EeV] & ($N=353$)\\
\hline
1 & $-2.19$ & 1.35 & 3.54 & $>60$ & 572 & $-1.67$ & 1.42 & 3.36 & 2.21 & 660 \\
2 & \phantom{-}0.16 & 5.75 & 3.65 & $>52$ & 605  & \phantom{-}0.51 & 5.96 & 3.53 & $>27$ & 661\\
3 & \phantom{-}0.56 & 7.41 & 3.75 & $>41$ & 651 & \phantom{-}0.81 & 7.49 & 3.64 & $>29$ & 699 \\

\end{tabular}}

\caption{Parameters of the fit to the spectrum and $X_{\rm max}$ distributions in the absence of a magnetic horizon effect, for the EPOS-LHC and Sibyll~2.3d hadronic interaction models and no source evolution, with the corresponding deviance $D$ and number of fitted data points $N$. Cutoff shapes with $\Delta =1$, 2 and 3 are considered.}
\label{table_0}
\end{table}

\subsubsection{Results obtained including the magnetic horizon effect}

When including the magnetic horizon effect, two new parameters enter in the fit, the normalized intersource distance $X_{\rm s}$ and the critical rigidity $R_{\rm crit}$. Note that since for intersource distances much smaller than the diffusion length in the EGMF the spectrum of the particles reaching Earth is not modified (in agreement with the propagation theorem \cite{al04}), in the limit $X_{\rm s} \rightarrow 0$ the results of the fit should coincide with the ones in the absence of magnetic fields presented above.

\label{sec_result_MHE}
\begin{figure}

  \hspace{0.5 cm}  \centering      \includegraphics[width=0.6\textwidth]{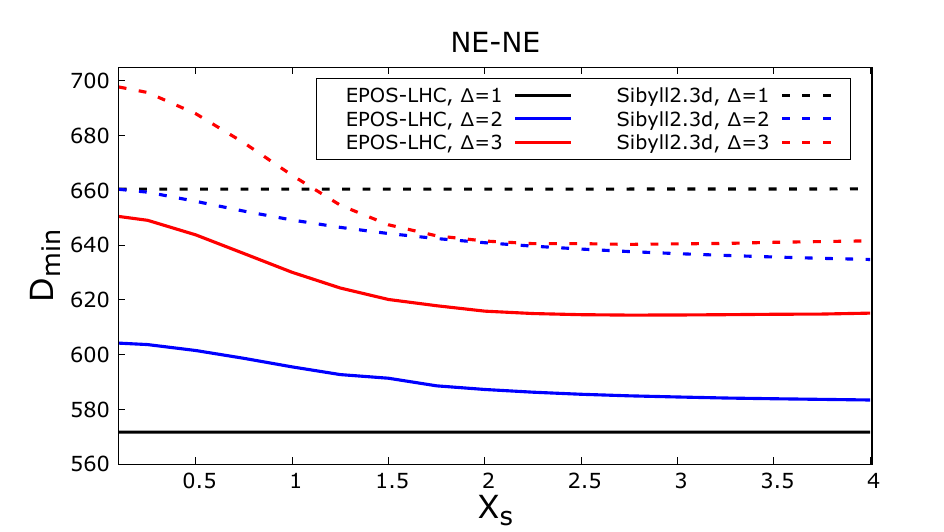}

\includegraphics[width=0.4\textwidth]{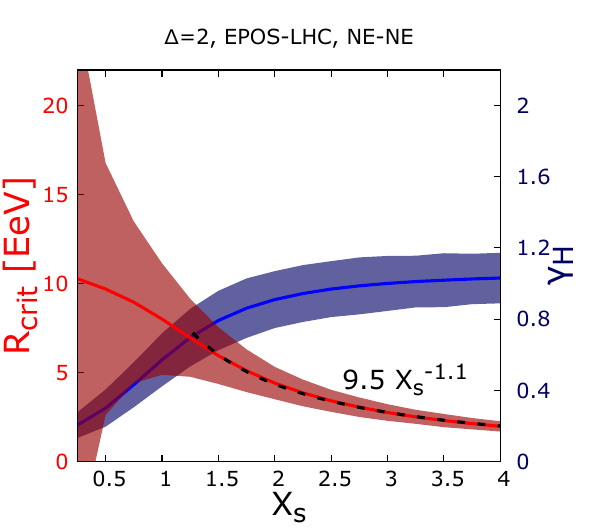} \ \ 
    \includegraphics[width=0.4\textwidth]{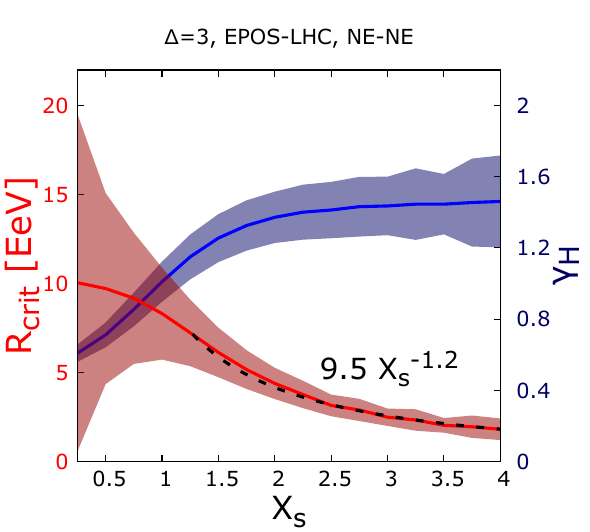}
   
  \hspace{0.5 cm}  
  
  \includegraphics[width=0.4\textwidth]{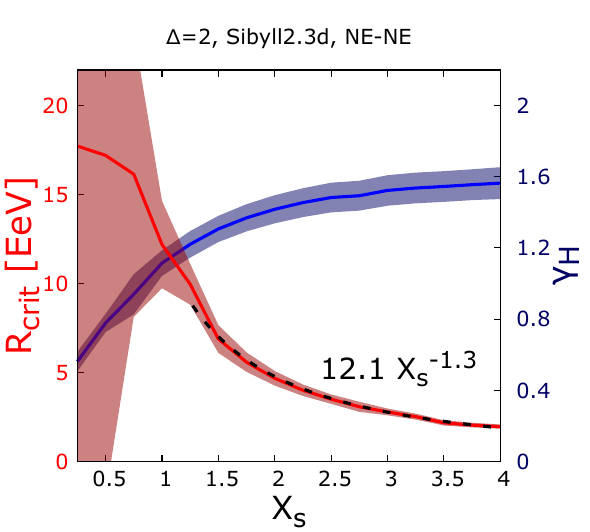} \ \ 
    \includegraphics[width=0.4\textwidth]{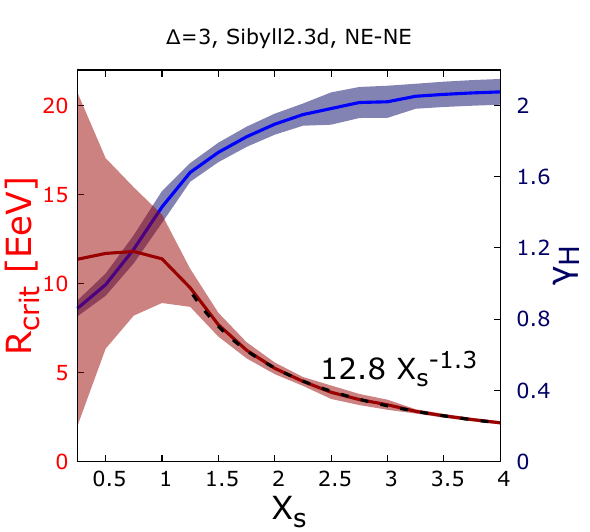}
  
    \caption{The top panel displays the  deviances as a function of $X_{\rm s}$ for the different scenarios of the  NE--NE case (i.e. NE for the LE and HE components). Lower panels display the  values of  $R_{\rm crit}$ and $\gamma_{\rm H}$ as a function of  $X_{\rm s}$ for the two hadronic models and for  $\Delta=2$ (left panels) or $\Delta=3$ (right panels). Also shown are analytic fits to $R_{\rm crit}$ as a function of $X_{\rm s}$. 
        }
    \label{fig:rcgamvsxs}
\end{figure}

To gain insight on the dependence of the results with the normalised inter-source distance parameter $X_{\rm s}$, we first present some results fixing different $X_{\rm s}$ values and fitting the rest of the parameters. 
We display in Fig.~\ref{fig:rcgamvsxs} (top panel) the deviances as a function of $X_{\rm s}$   for  the different  cases considered of  non-evolving sources, i.e. for the two hadronic models and for the  cutoff functions with $\Delta=1$, 2 and 3.  One can see that in the case with $\Delta=1$ the deviance is degenerate for all $X_{\rm s}$ values, since the fit  actually favours the no magnetic field case (i.e. being compatible with a vanishing $R_{\rm crit}$),  and preferring a negative $\gamma_{\rm H}$ for all values of $X_{\rm s}$. For the cases  with $\Delta=2$ and 3 the fit actually favours the case where the magnetic horizon effect plays a significant role, since smaller deviances are obtained for larger $X_{\rm s}$, and the deviance becomes almost degenerate for $X_{\rm s} \geq 2$. 
The  smallest deviance is obtained for $\Delta=1$ (no magnetic horizon), but let us  mention that when including the possible impact of systematic experimental uncertainties, the models leading to the smallest deviance may change, as will be discussed in the next Section.
In the four lower panels of the  figure we show with blue bands the best fitting spectral index of the HE population and with red bands the critical rigidity as a function of $X_{\rm s}$ for the cases  with $\Delta=2$ and 3, for which the presence of an EGMF  is preferred.
Here softer associated values of the spectral indices $\gamma_{\rm H}$ result for increasing $X_{\rm s}$  as a consequence of the magnetic horizon effect. 
It is interesting to note that larger normalised distances also give rise to smaller best-fitting critical rigidities, since smaller magnetic field strengths are required to produce the spectral suppression at low energies when the sources are farther apart, as it was shown in Fig.~\ref{fig:Gfactor}.
We also show in the plots (analytic dashed lines) the approximate dependence of $R_{\rm crit}$ as a function of $X_{\rm s}$ in the region where the magnetic horizon effect plays a relevant role. In the four cases (EPOS-LHC and Sibyll~2.3d with $\Delta = 2$ or 3),  the best fits are obtained for models satisfying $X_{\rm s} R_{\rm crit} \simeq 10$\,EeV.

\begin{table}[ht]
\centering
{\small
\scalebox{0.83}{
\begin{tabular}[H]{ @{}c| c c c c c c c| c c c c c c c @{}}

 \multicolumn{14}{c}{with EGMF, NE-NE}
\\
\hline
& \multicolumn{7}{c|}{EPOS-LHC} & \multicolumn{7}{c}{Sibyll 2.3d}  \\
\hline
$\Delta$ & $\gamma_{\rm H}$ & $R_{\rm cut}^{\rm H}$ & $\gamma_{\rm L}$ & $R_{\rm cut}^{\rm L}$ & $X_{\rm s}$ & $R_{\rm crit}$ & $D$& $\gamma_{\rm H}$ & $R_{\rm cut}^{\rm H}$ & $\gamma_{\rm L}$ & $R_{\rm cut}^{\rm L}$ &  $X_{\rm s}$ & $R_{\rm crit}$ & $D$\bigstrut[t]\\ 
& &  [EeV] & &  [EeV] &  & [EeV] & $(N=353)$ & &  [EeV] & &  [EeV] & & [EeV] & $(N=353)$\\
\hline
1 &  $-2.19$ & 1.35 & 3.54 & $>60$ & 0\phantom{.0} & -- & 572 & $-1.67$ & 1.42 & 3.37 & 2.21 & 0\phantom{.0} & -- & 660 \\
2 & \phantom{-}1.03 & 6.02 & 3.62 & $>51$ & $>3.2$ & 1.97 & 583 & \phantom{-}1.35 & 6.22 & 3.53 & $>25$ & $>3.1$ & 1.54 & 635 \\
3 & \phantom{-}1.43 & 7.50 & 3.69 & $>61$ & 2.8 & 2.79 & 614 & \phantom{-}2\phantom{.00} & 7.50 & 3.62 & $>31$ & 2.6 & 3.77 & 640 \\
\hline
\multicolumn{7}{l}{  }&\multicolumn{7}{l}{SFR-NE} \\
\hline
1 & $-2.09$ & 1.39 & 3.24 & $>63$ & 0\phantom{.0} & -- & 578 & $-1.64$ & 1.44 & 3.03 & 2.89 & 0\phantom{.0} & -- & 665 \\
2 & \phantom{-}1.12 & 6.14 & 3.33 & $>61$ & $>3.5$ & 2.11 & 586 & \phantom{-}1.45 & 6.29 & 3.21 & $>37$ & $>3.2$ & 1.67 & 635 \\
3 & \phantom{-}1.49 & 7.52 & 3.41 & $>57$ & 2.7 & 3.15 & 617 & \phantom{-}2.07 & 7.49 & 3.31 & $>33$ & 2.8 & 3.52 & 637\\
\end{tabular}}}

\caption{Parameters of the fit to the flux and composition including the magnetic horizon effect for the EPOS-LHC and Sibyll~2.3d hadronic interaction models and for different cosmological evolutions of the low-energy component and NE for the high-energy  component. Cutoff shapes with $\Delta =1$, 2 and 3 are considered. }
\label{table_1}
\end{table}

In Table \ref{table_1} the best fitting values of the spectral and magnetic suppression parameters, obtained by minimizing also on $X_{\rm s}$ within the range [0, 4],  are reported.  For EPOS-LHC the minimum deviance corresponds to the $\Delta=1$ cutoff case (for which the fit is equivalent to the no magnetic field case), but for Sibyll~2.3d the deviance is smaller for the steeper cutoffs. For $\Delta =2$ and 3 a significantly
softer  spectrum of the HE population  with respect to the case with no magnetic field is obtained, arising from the effect of the magnetic horizon which significantly hardens the spectrum of the CRs arriving to the Earth and hence allows for softer spectra at the sources. The spectral indices obtained in this case are larger than unity, and in particular for Sibyll~2.3d with  $\Delta=3$ a value $\gamma_{\rm H}\simeq 2$ is obtained, which is well within the expectations from diffusive shock acceleration.

\begin{figure}[b]
    \centering
\includegraphics[width=0.49\textwidth]{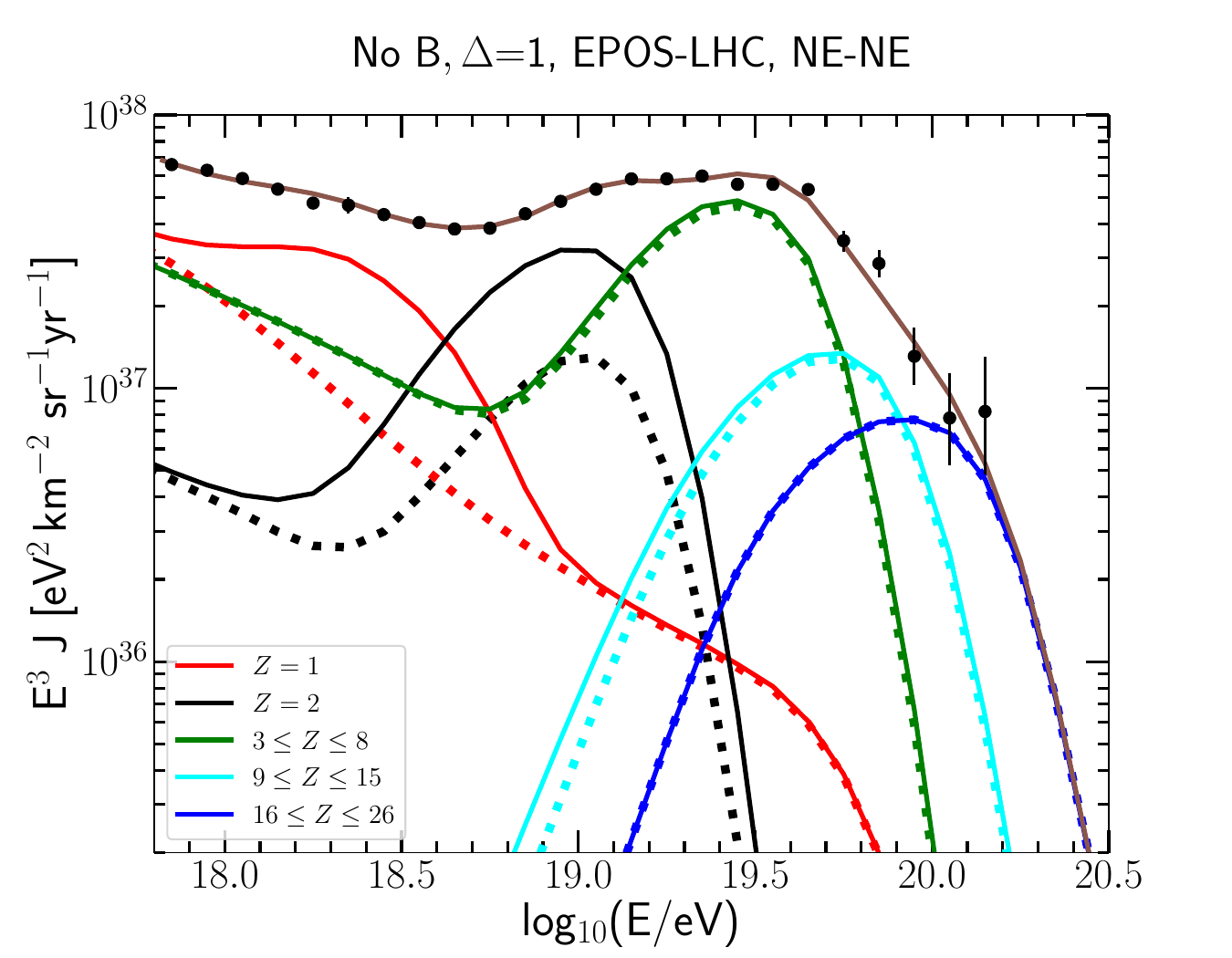}
    \includegraphics[width=0.49\textwidth]{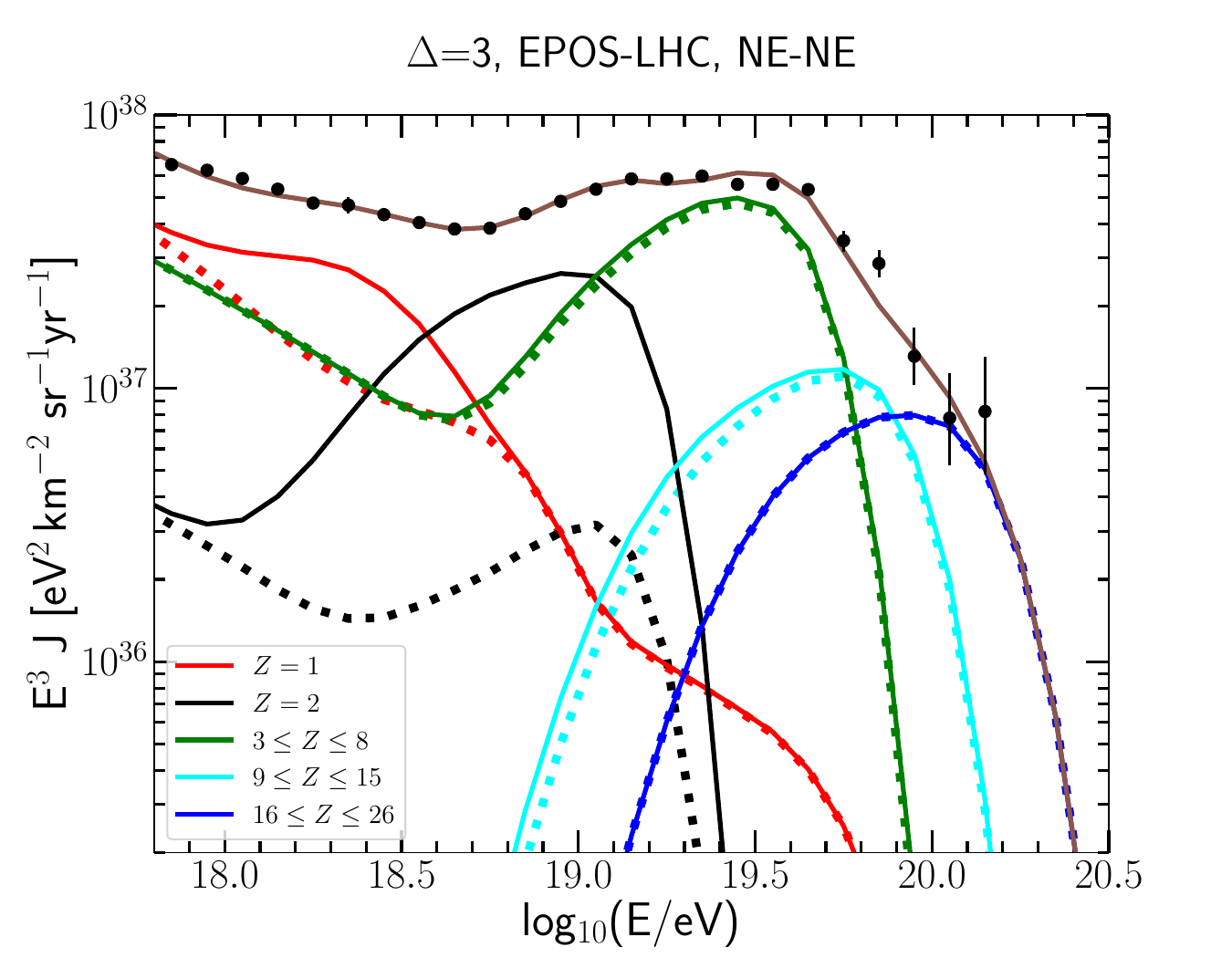}

    \includegraphics[width=0.49\textwidth]{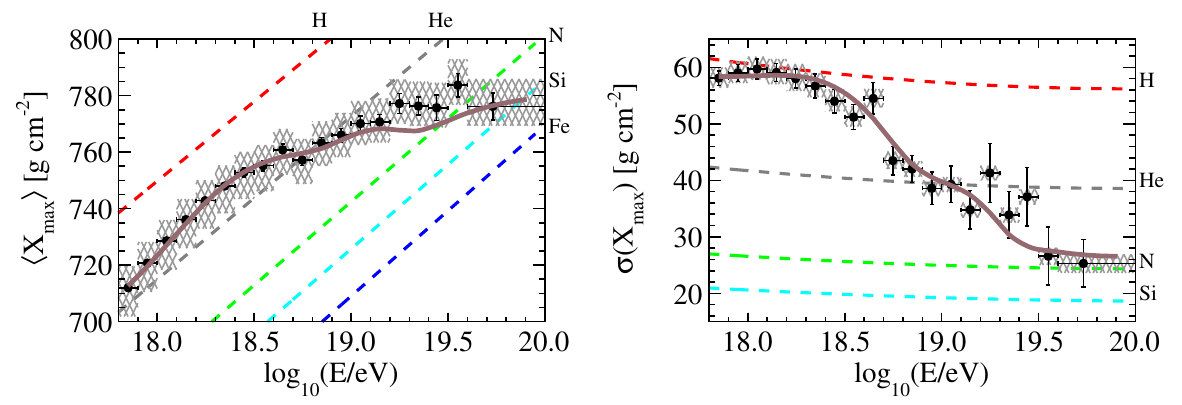}
    \includegraphics[width=0.49\textwidth]{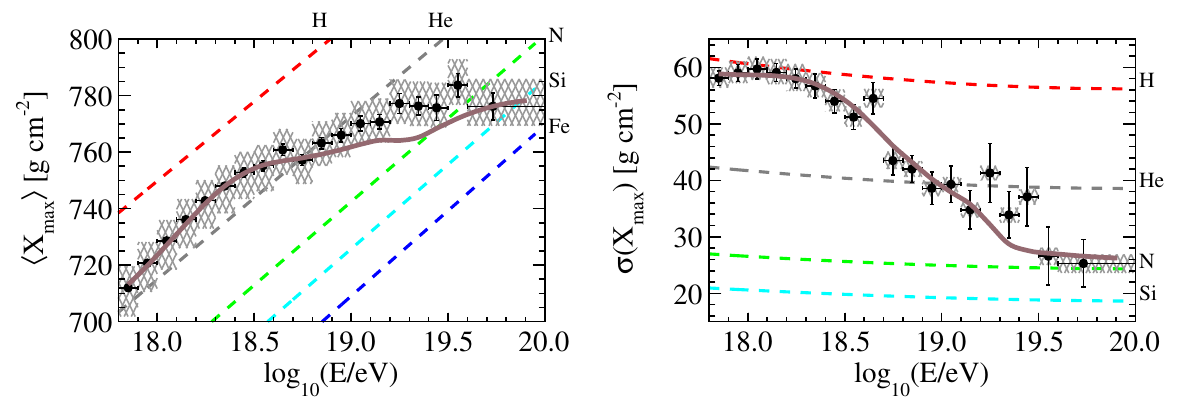}
    
    \caption{Flux at Earth (upper panels) and moments of the $X_{\rm max}$ distribution (lower panels) for the best fit models, adopting the EPOS-LHC hadronic model. All the scenarios assume NE-NE for the cosmological evolution of the sources. Dotted lines in the spectrum plots represent the flux coming from the primary nuclei while solid lines correspond to the total flux (primary  plus secondary) of each mass group. The left panel depict the results for the $\Delta = 1$ cutoff case, for which the fit prefers the no magnetic horizon solution, while the right panel depict the $\Delta = 3$ cutoff case, where the best fit has a significant magnetic horizon.}
    \label{fig:fitepos}
\end{figure}

We also report in Table   \ref{table_1} 
the results of the fit when considering a scenario involving SFR evolution for the LE component and NE for the HE component (labelled as SFR-NE). This scenario leads to slightly harder LE spectrum, with $\gamma_{\rm L}$ being  smaller by about 0.3, due to the enhanced emission at high redshifts that leads to more significant steepening of the spectrum at the Earth than the NE case due to propagation effects. The rest of the parameters and the deviances are very similar to the NE-NE case. Considering  scenarios with SFR evolution for the high-energy population leads always to worse  deviances (in agreement with \cite{xcf}) and thus they will not be further discussed here.

\begin{figure}[t]
    \centering

    \includegraphics[width=0.49\textwidth]{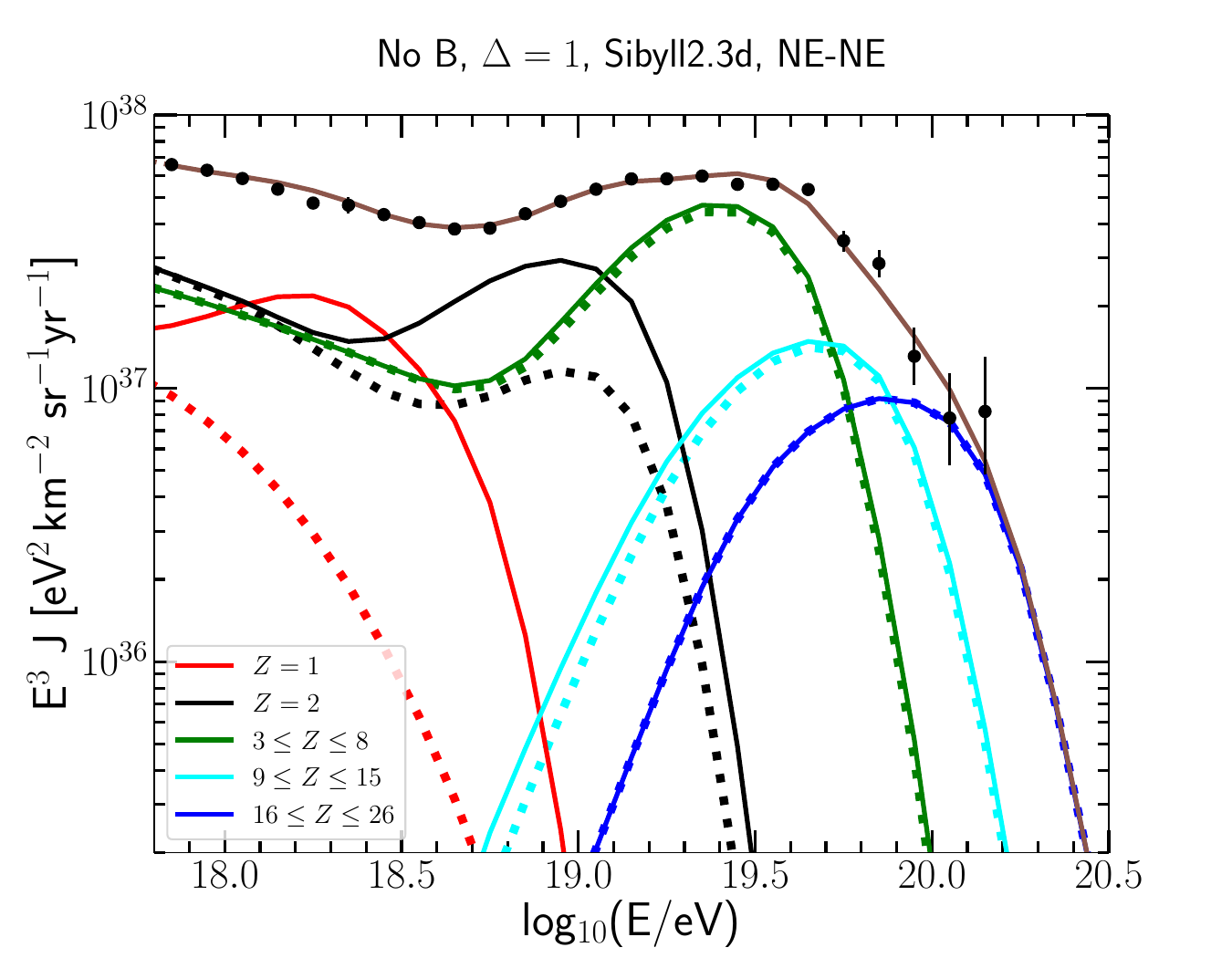}
    \includegraphics[width=0.49\textwidth]{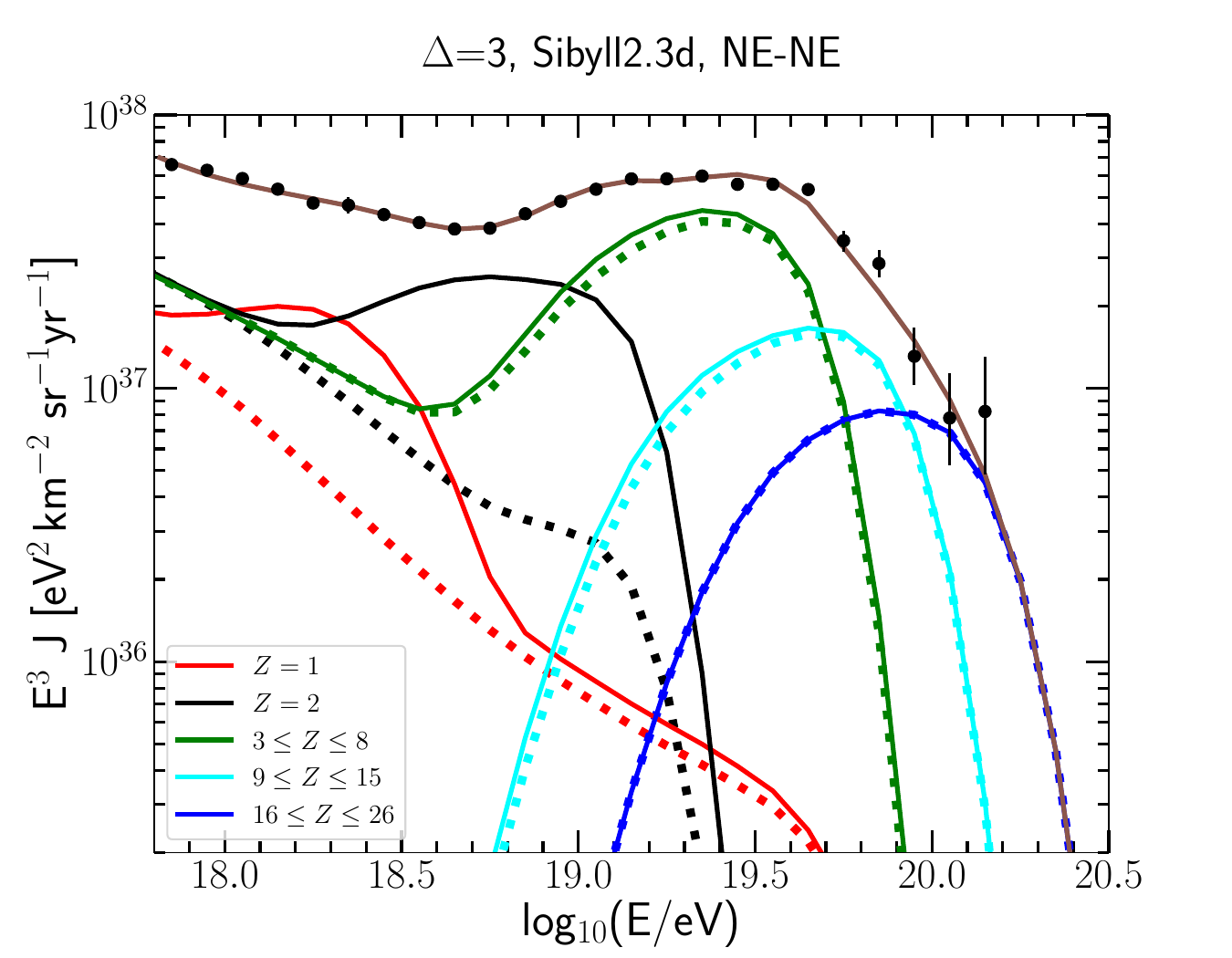}

    \includegraphics[width=0.49\textwidth]{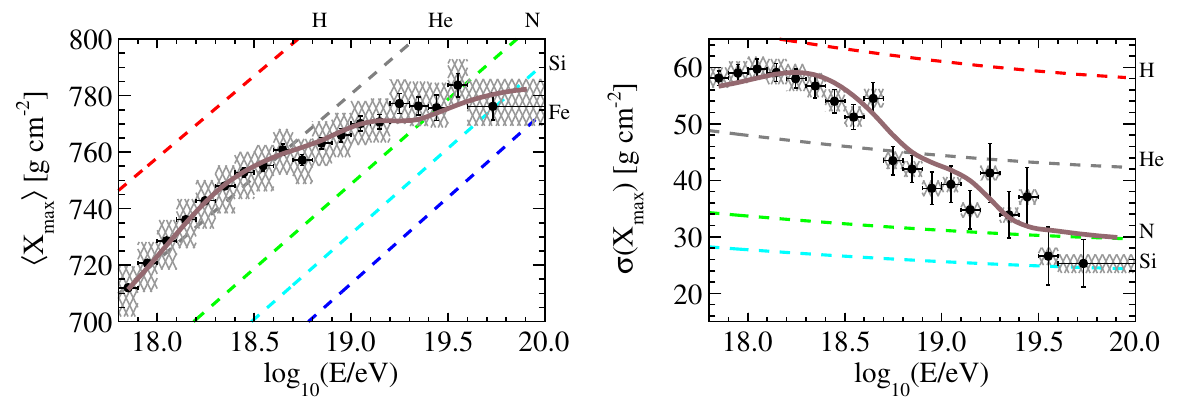}
    \includegraphics[width=0.49\textwidth]{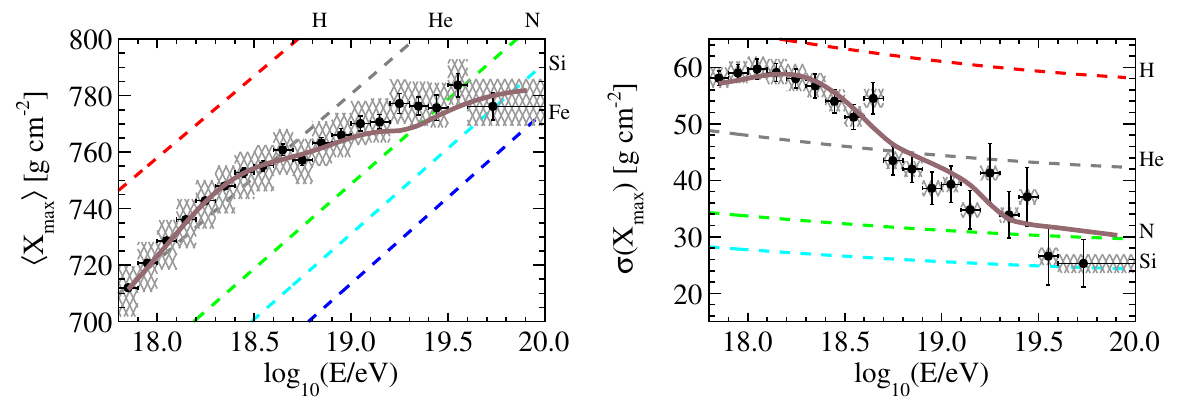}
    
    \caption{Analogous to Fig.~\ref{fig:fitepos}, but for the Sibyll~2.3d hadronic interaction model.}
    \label{fig:fitsibyll}
\end{figure}

Figures \ref{fig:fitepos} and \ref{fig:fitsibyll} present the spectrum at Earth and the first two $X_{\rm max}$ moments for EPOS-LHC and Sibyll~2.3d, respectively. The left panels show the results for a cutoff with $\Delta =1$, for which a significant magnetic horizon effect is not favoured by the fit. The right plots are for $\Delta=3$, where the inclusion of the magnetic suppression leads to a better fit. The case with $\Delta =2$ is qualitatively similar to that with $\Delta =3$. Although the full $X_{\rm max}$ distribution was fitted, as explained in Section \ref{fit}, we display for illustration of the goodness of the fit the results for the first two $X_{\rm max}$ moments. 

\begin{figure}[t]
    \centering

    \includegraphics[width=0.48\textwidth]{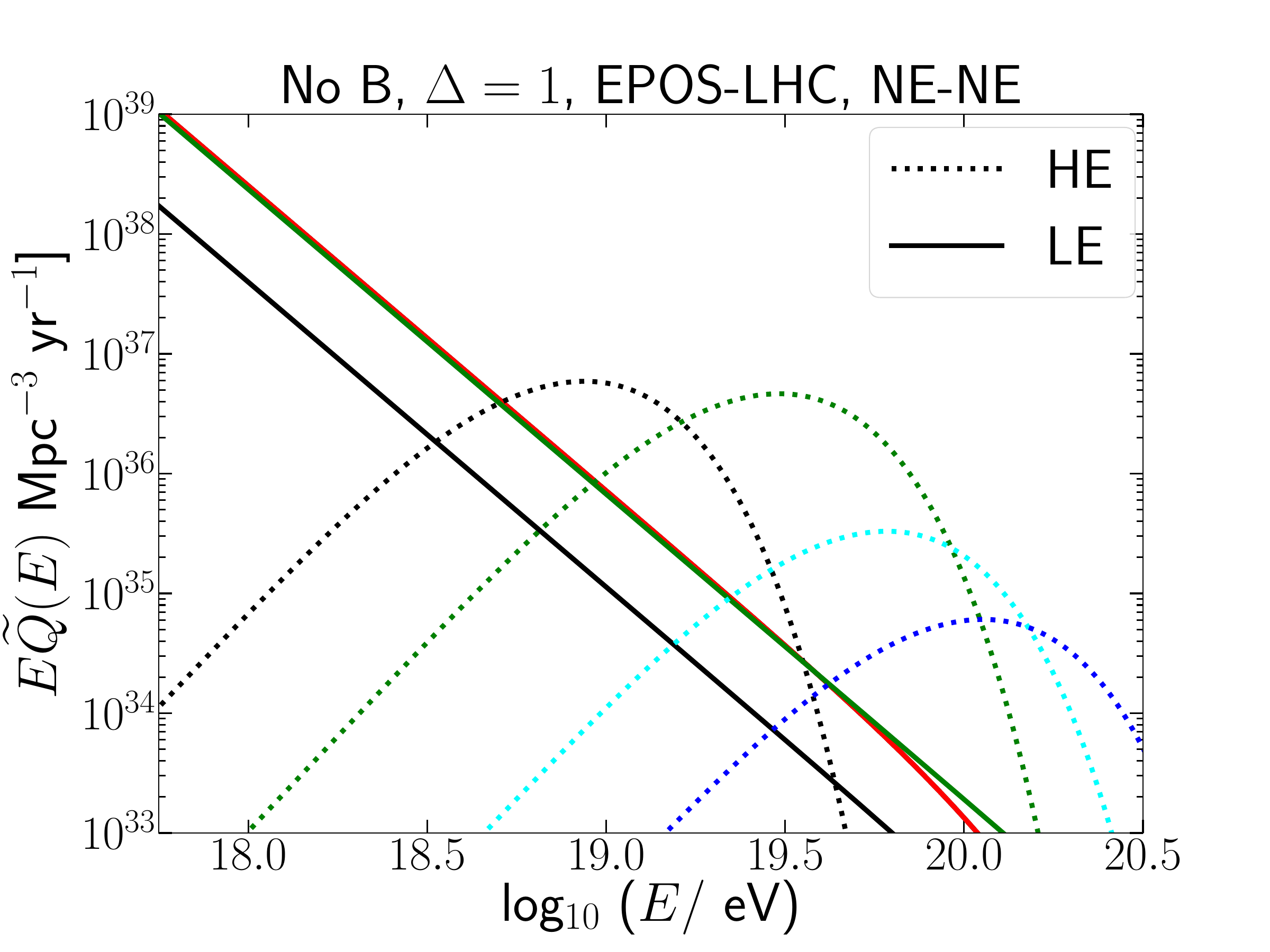}
    \includegraphics[width=0.48\textwidth]{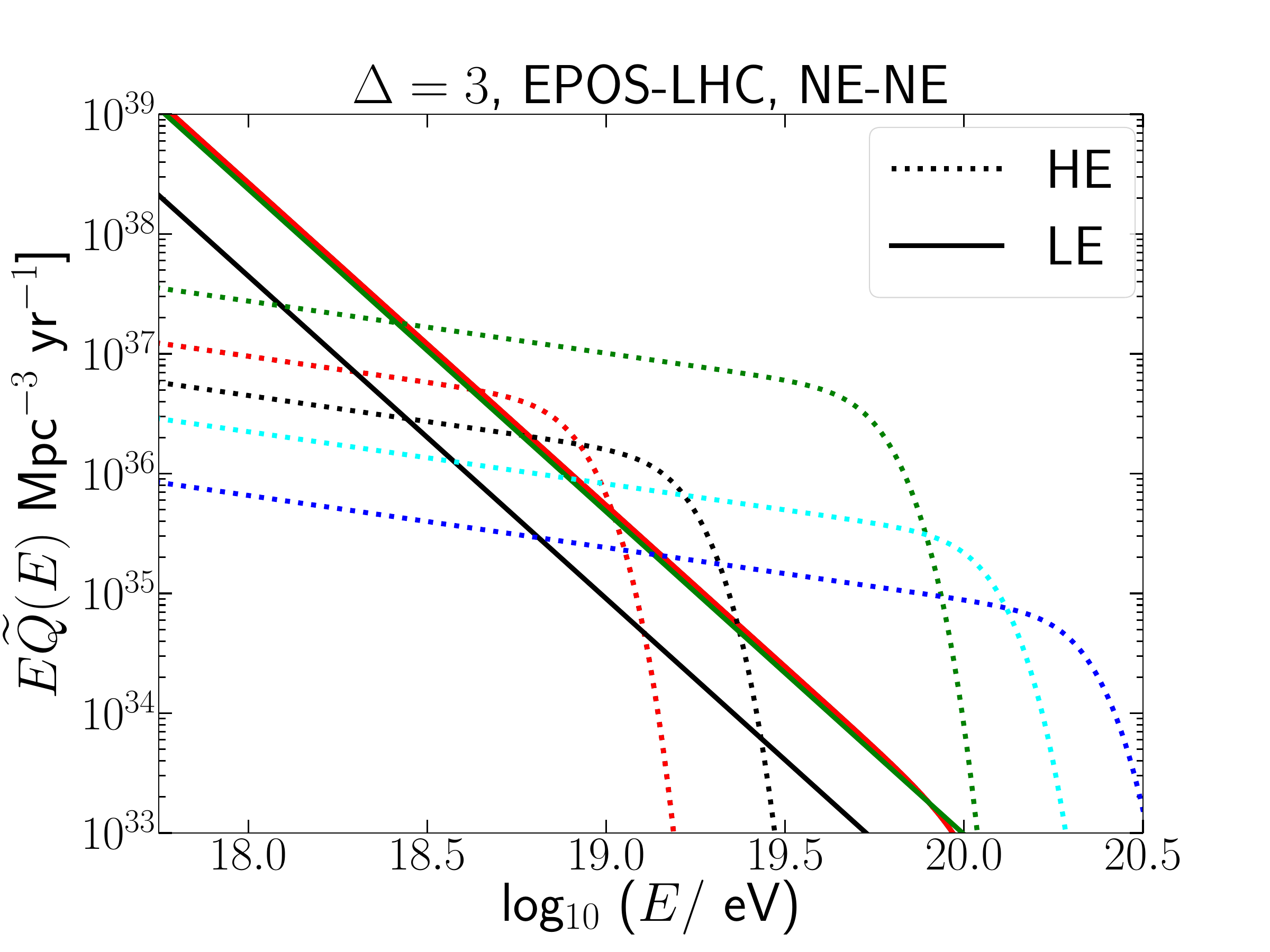}
    \includegraphics[width=0.48\textwidth]{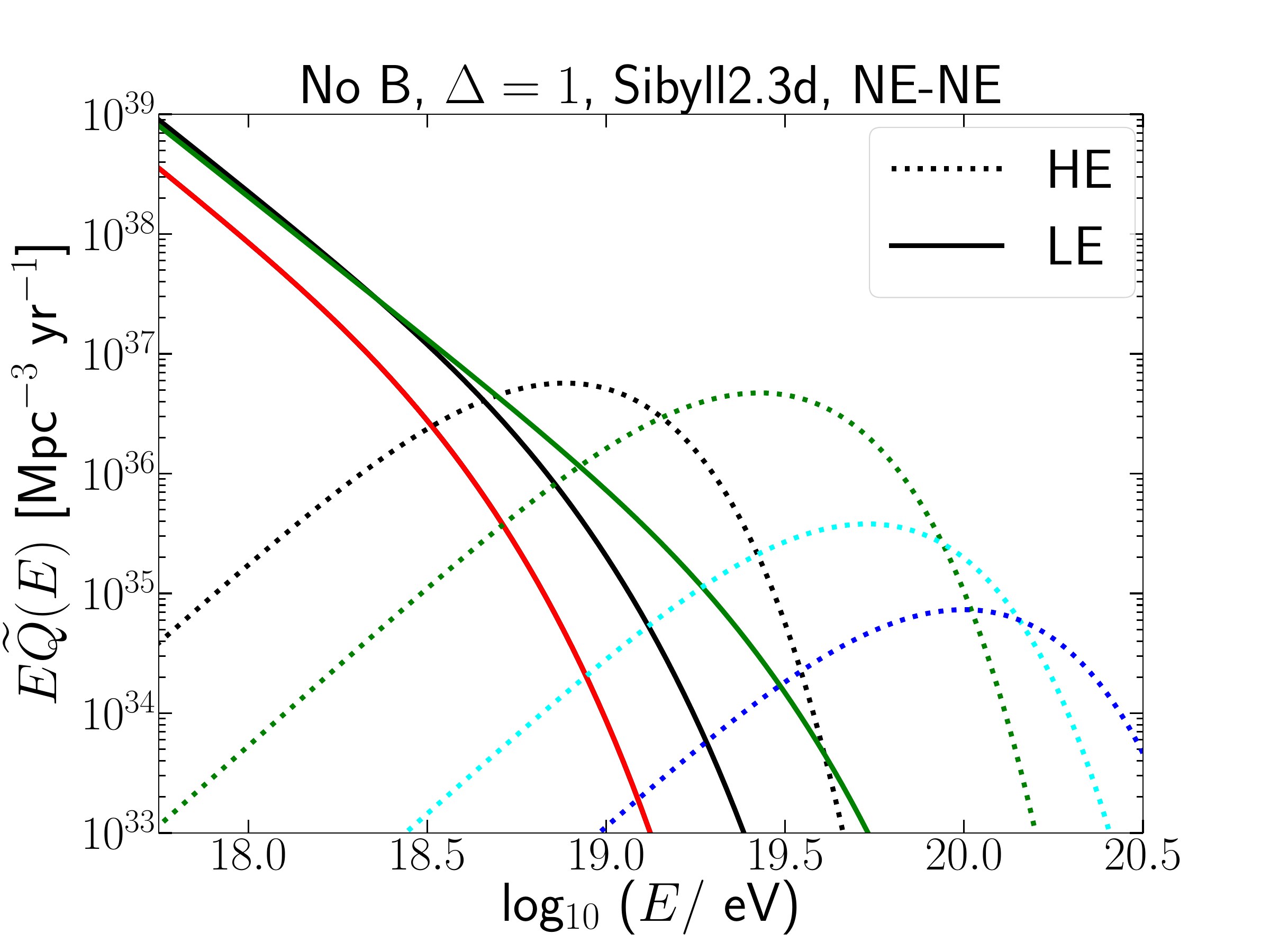}     \includegraphics[width=0.48\textwidth]{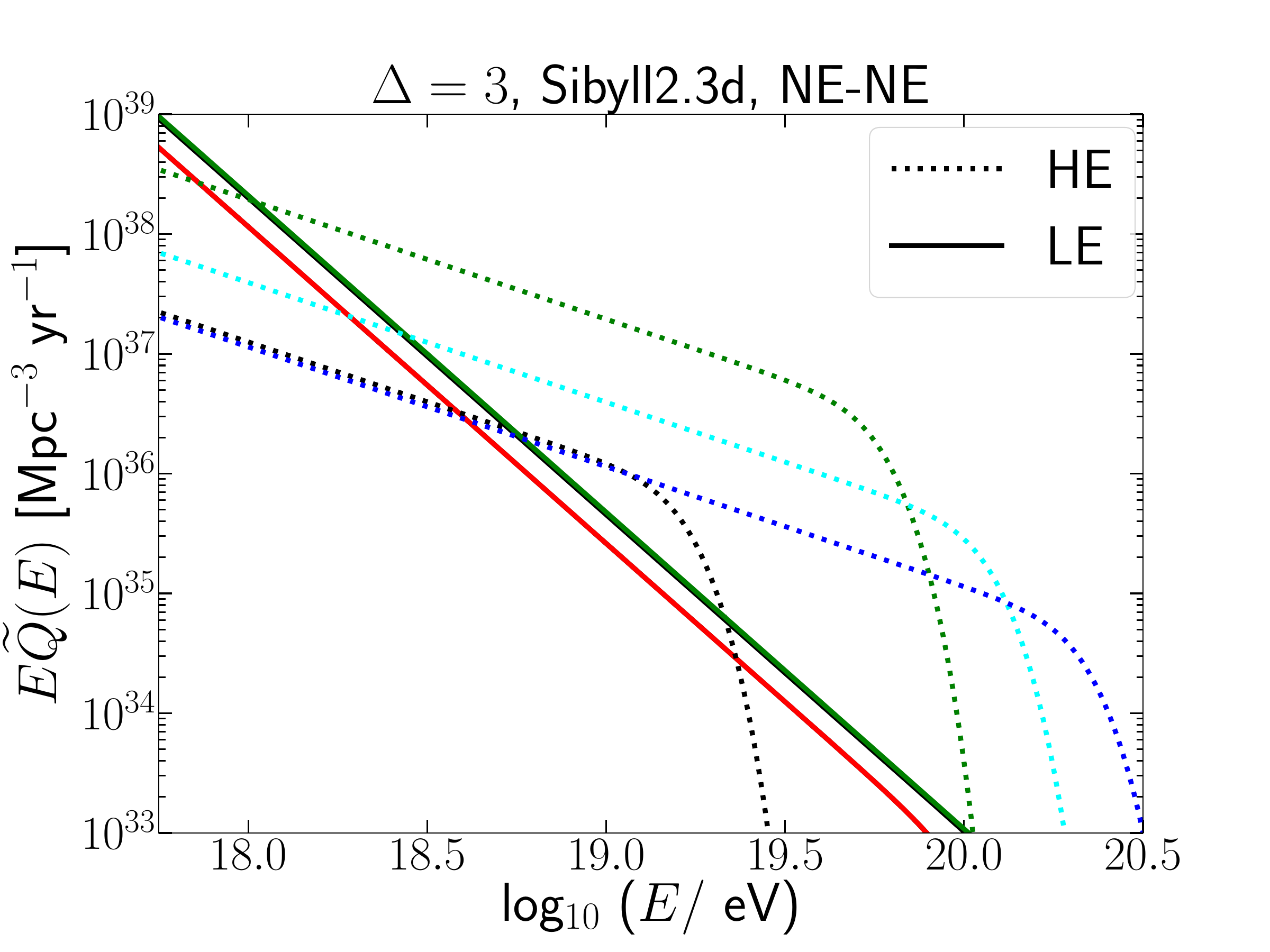}
     \caption{Differential injection rate of particles at the source per logarithmic energy bin for H (red), He (gray), N  (green), Si   (cyan) and Fe  (blue) for the scenarios considered in Figs. \ref{fig:fitepos} and \ref{fig:fitsibyll}: the EPOS-LHC and Sibyll 2.3d hadronic interaction models and the $\Delta = 1$ and $\Delta = 3$ cutoff shapes for NE of the source luminosities.  }
    \label{fig:injection_B}
\end{figure}

Figure \ref{fig:injection_B}  shows  the differential particle generation rate per logarithmic energy bin of each component at the sources for the different scenarios discussed above. The LE population component, depicted in solid lines, show a soft spectrum for the cases. Note in particular the difference in the spectrum of the HE component (in dotted lines) between the left panels corresponding to $\Delta =1$, where there is no magnetic horizon for the best fit and the spectrum at the source is very hard, and the right panels corresponding to $\Delta = 3$, where the magnetic horizon is significant and the spectrum is much softer.

Some features which can be inferred from the results of the fits are:
\begin{itemize}
    \item  In all  cases  the lighter (H and He) nuclei reaching the Earth at energies of few EeV are actually mostly secondaries, arising mainly from the disintegration of N nuclei from the high-energy population. 

    \item There is a significant amount of secondary protons  around 1--3~EeV, which leads to a relatively light composition in this energy range.

    \item The He nuclei  are mainly responsible for the instep feature (the change in slope observed at about 15~EeV), whose energy is  to some extent related to the energy at which He nuclei get strongly disintegrated by CMB photons but is also affected by the source cutoff suppression. However, in the case of Sibyll~2.3d with $\Delta=3$ the N contribution is more significant at the instep and is also partially responsible for this feature. 
    
    \item Beyond the instep feature, the spectrum is largely dominated by the N component. The suppression above 50~EeV is in part due to the attenuation of the N flux by interactions with the CMB and in part due to the cutoff of the high-energy population,  with the cutoff suppression of the N nuclei appearing at an energy $\sim 7/2=3.5$ times larger than the He one. At the highest energies the  Si and Fe  elements give the main contribution to the flux. Note that above $10^{19.6}$\,eV there is just one integral energy bin of $X_{\rm max}$, so that  the composition information here is quite limited, and improvements in this respect are expected to be obtained with the use of the surface detector information and the recent upgrade of the Observatory \cite{upgrade}.

    \item Regarding the low-energy component, most of the flux at EeV energies  arises from the H and N components, with a sizeable contribution also from He in the case of Sibyll~2.3d, and no significant amounts of Si or Fe.

    \item In the cases where the cutoff rigidity of the low-energy component slides to the limit allowed in the fit ($R^{\rm L}_{\rm cut}=100$~EeV) the deviance is actually almost unchanged for values larger than $\sim 25$~EeV. These large cutoff values give rise to a subdominant component of H nuclei extending up to very high energies, which could have implications for the production of cosmogenic neutrinos up to EeV energies \cite{xcf}. We also note that often a secondary minimum of the deviance exist with a lower cutoff rigidity $R^{\rm L}_{\rm cut}\simeq 3$~EeV and a deviance in general larger by a few units, as discussed in \cite{xcf}, although this minimum is actually the global one for Sibyll~2.3d and $\Delta=1$.

\item The EPOS-LHC hadronic model generally predicts that showers of a given composition and energy are less penetrating (have smaller $X_{\rm max}$) than those from the Sibyll~2.3d model. This leads to a lighter inferred composition when the EPOS-LHC model is considered, as is apparent  in the fact that  for this model there is an enhanced contribution of H observed below the ankle, the He contribution is more prominent around the instep and a slightly smaller Si contribution is present in the suppression region.
    
\end{itemize}

We report in Table \ref{table_2} the results of the minimization for the full set of  fitted parameters, including their  statistical uncertainties,  for the  scenarios shown in Figs.~\ref{fig:fitepos} and \ref{fig:fitsibyll}. They correspond to one case where the magnetic horizon effect plays an important role ($\Delta=3$) and another where it does not ($\Delta=1$), both for the EPOS-LHC and Sibyll~2.3d hadronic models. 
We also quote the present day emissivities above a threshold energy $E_{\rm th}=10^{17.8}$~eV, $L_{44}^a\equiv\sum_AL^a_A/(10^{44}\,\rm{ erg\, Mpc^{-3}\, yr^{-1}})$, where
\begin{equation}
     L^a_A\equiv  \int_{E_{\rm th}}^\infty {\rm d}E\,E\tilde Q^a_A(E, z=0).
    \label{eq_lum}
\end{equation}
It is worth noting that the emissivities inferred for the HE population in the scenarios where the magnetic horizon effect plays a significant role are larger than in the absence of EGMF, since in the first case there is a sizeable emission at low energies which doesn't manage to arrive to the Earth.

\begin{table}[t]
\centering

{\small
\begin{tabular}[H]{ @{}c| c  c |c c @{}}
NE-NE&\multicolumn{2}{c|}{EPOS-LHC} & \multicolumn{2}{c}{Sibyll 2.3d} \\
\hline
$\Delta$ &  1 &  3 & 1 & 3 \\
\hline
$X_{\rm s}$ & --- &  $2.8 _{-0.7}^{+1.2}$ & --- & $2.6_{-0.5}^{+1.1}$ \bigstrut[t]\\
$R_{\rm crit} [\rm{EeV}]$ &  --- &  $2.8 \pm 1.5 $ & --- & $3.8 \pm 1.4$ \\
\hline
High energy&\\
\hline
$\gamma_{\rm H}$ & $-2.19 \pm 0.10$ & \phantom{+}1.43$^{+0.16}_{-0.22}$ & $-1.67\pm 0.13$ & \phantom{+}2.00$^{+0.10}_{-0.11}$  \bigstrut[t]\\
$R_{\rm cut}^{\rm H}$ [EeV] & $1.35 \pm{0.04}$ & $7.50\pm 0.15$ & $1.42\pm 0.05$ & \phantom{+}7.50$_{-0.20}^{+0.18}$ \\
$f_{\rm H} $ & $<0.1$ & $21 \pm 11 $ & $<10^{-3}$ & $<10^{-2}$\\

$f_{\rm He}$ & $98.6_{-0.2}^{+0.1}$ & $10.1 \pm 5.9$ & $97.1\pm 0.6$ & \phantom{0}$5.0\pm 5.0 $\\
\ \ \ \ \ \ $f_{\rm N}$\  [\%]& \phantom{+}$1.4_{-0.5}^{+0.3}$ & $61.9_{-10.4}^{+8.8}$ & \phantom{+}$2.8_{-0.6}^{+0.7}$ & $75.4_{-9.7}^{+9.1}$  \\
$f_{\rm Si}$ & \phantom{0}$<10^{-3}$ & \phantom{0}$5.0_{-2.4}^{+2.7}$ & \phantom{+}$<10^{-2}$ & $15.2_{-5.4}^{+4.7}$\\
$f_{\rm Fe}$ & \phantom{0}$<10^{-4}$ & \phantom{0}$1.5_{-0.7}^{+0.9}$ & \phantom{0}$<10^{-3}$ & \phantom{0}$4.4_{-1.9}^{+1.7}$ \\
$L_{44}^{\rm H}$ & $5.0 \pm 0.1 $ & $9.3 \pm 2.6$ & $4.9\pm 0.1$ & $18.4 \pm 2.9$ \\
\hline

Low energy&\\
\hline
$\gamma_{\rm L}$ & \phantom{00}3.54$\pm 0.03$ & \phantom{00}3.69$\pm 0.04$ & \phantom{+}3.37$_{-0.05}^{+0.04}$ & \phantom{00}3.62$\pm 0.04$ \bigstrut[t]\\
$R_{\rm cut}^{\rm L}$ [EeV] & $>60$ & $>49$ & \phantom{0}2.21$^{+0.55}_{-0.48}$ & $>30$  \\
$f_{\rm H}$ & $47.9 \pm 2.6$ & $51.7 \pm 2.3$ & $17.7\pm 2.5$ & $21.9 \pm 2.1$ \\
$f_{\rm He}$ & \phantom{+}$7.5 \pm 4.1$ & \phantom{0}$4.8 \pm 3.6 $ & $43.5_{-3.8}^{+3.6}$ & $38.1_{-3.7}^{+3.4}$\\
\ \ \ \ \ \ $f_{\rm N}$\  [\%] & $44.6_{-2.5}^{+2.2}$ & $43.4_{-2.5}^{+1.7}$ & $38.7\pm 2.0$ & $39.9 \pm 1.9$\\
$f_{\rm Si}$ & $<10^{-4}$ & $<10^{-2}$ & $<10^{-4}$ & $<10^{-7}$  \\
$f_{\rm Fe}$ & $<10^{-5}$ & $<10^{-2}$ & $<10^{-5}$ & $<10^{-4}$ \\
$L_{44}^{\rm L}$& $11.0\pm 0.2$ & $11.6 \pm 0.2$ & $10.8\pm 0.1 $ & $11.4 \pm 0.4 $ \\
\hline \hline 
$D\,  (N=353)$ & 572 & 614 & 660 & 640\\

\end{tabular}}

\caption{Parameters of the fit to the flux and composition for the scenarios shown in Figs.~\ref{fig:fitepos} and \ref{fig:fitsibyll}. We include for the EPOS-LHC and Sibyll~2.3d hadronic interaction models a scenario with  $\Delta=1$
(for which the magnetic horizon effect is not relevant), as well as  a scenario with $\Delta=3$ and  with EGMF for NR of source luminosities. Quoted are the fitted $X_{\rm s}$ and $R_{\rm crit}$, the spectrum shape parameters and element fractions for the two populations as well as the source emissivities above $10^{17.8}$~eV expressed in units of $10^{44}\,\rm{ erg\, Mpc^{-3}\, yr^{-1}}$.  The bottom row indicates the associated deviances and the number $N$ of data points considered. }
\label{table_2}
\end{table}

\subsection{Impact of systematic uncertainties on the energy and \texorpdfstring{$X_{\rm max}$}{Xmax} calibrations}

\begin{figure}[t]
    \centering
    \includegraphics[width=0.325\textwidth]{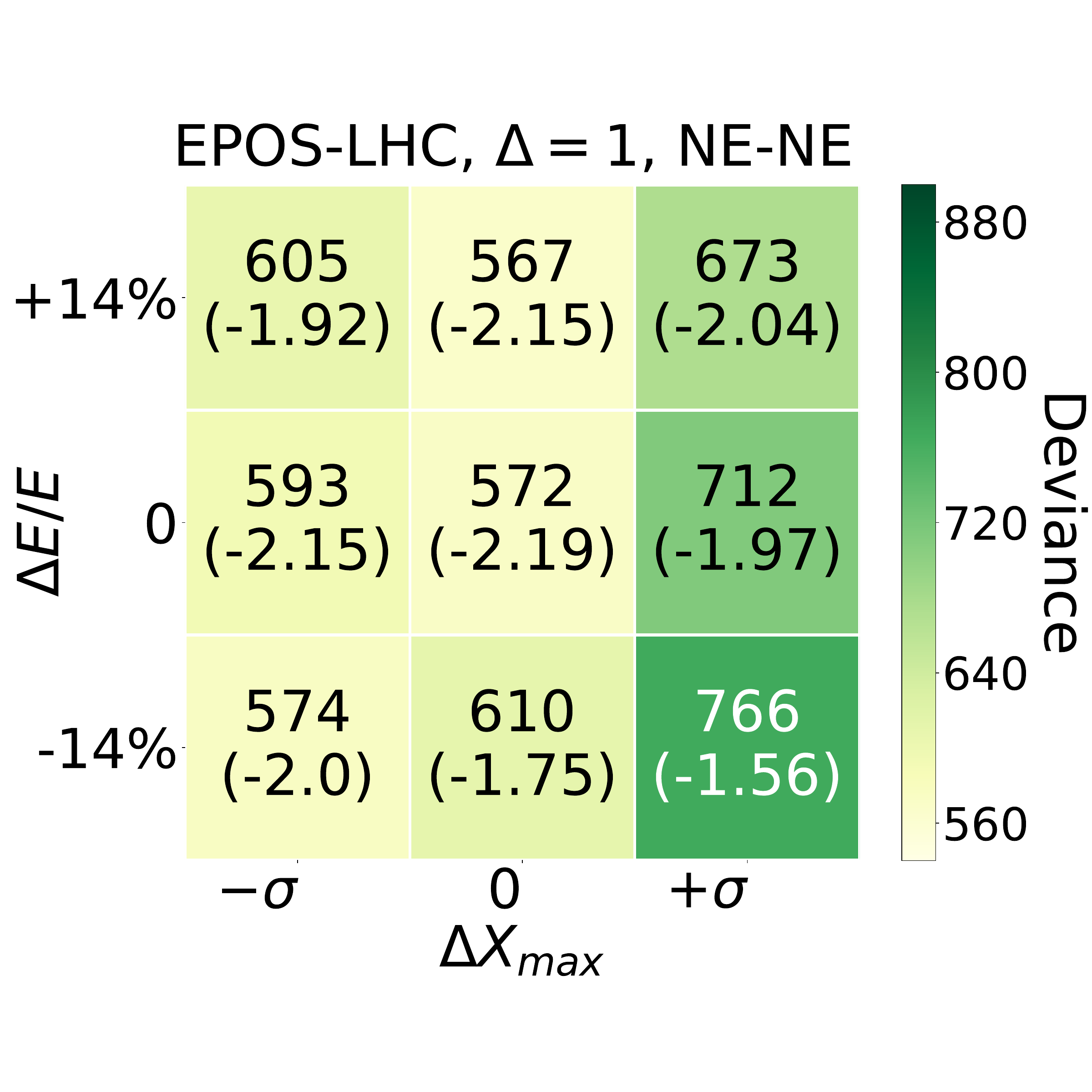}
    \includegraphics[width=0.325\textwidth]{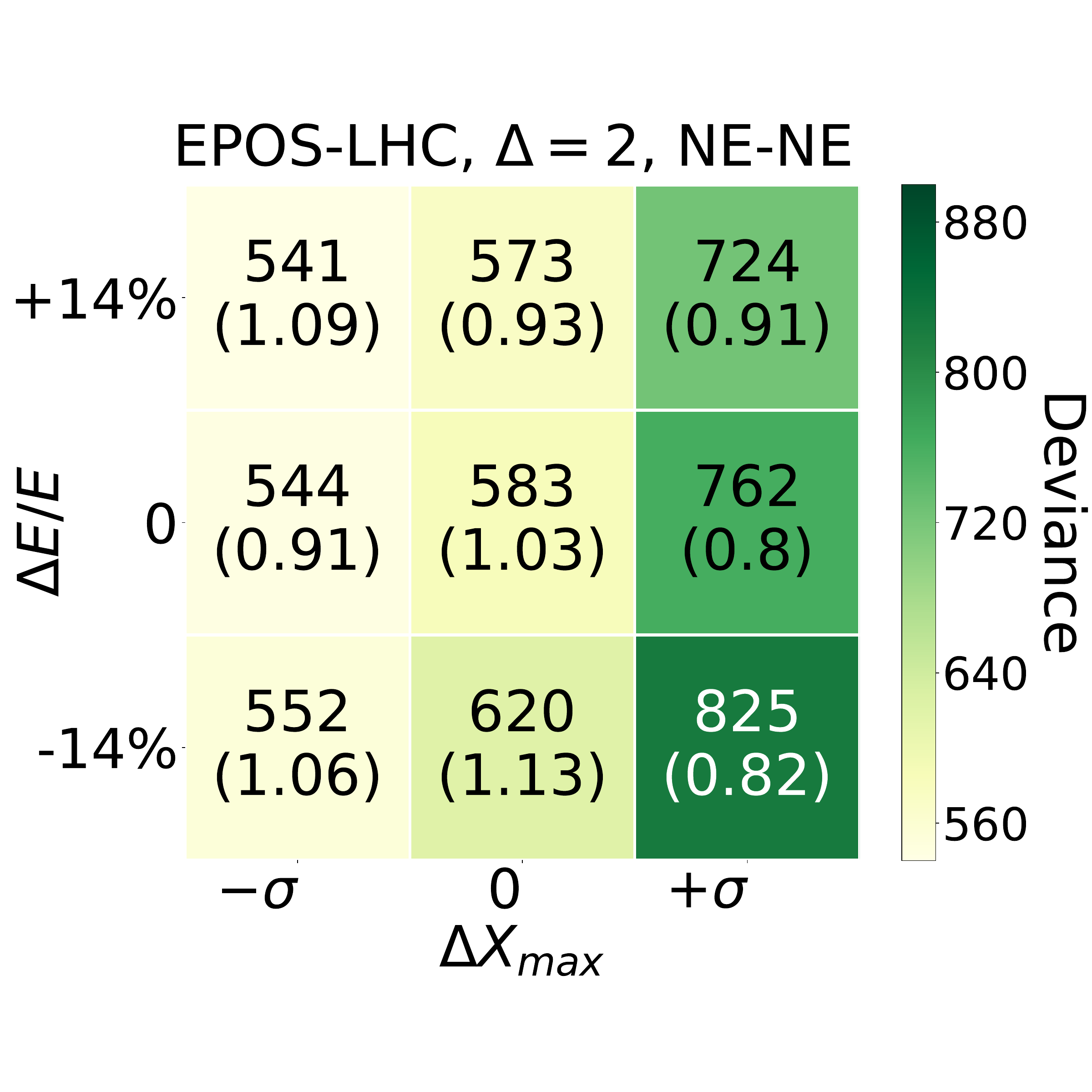}
    \includegraphics[width=0.325\textwidth]{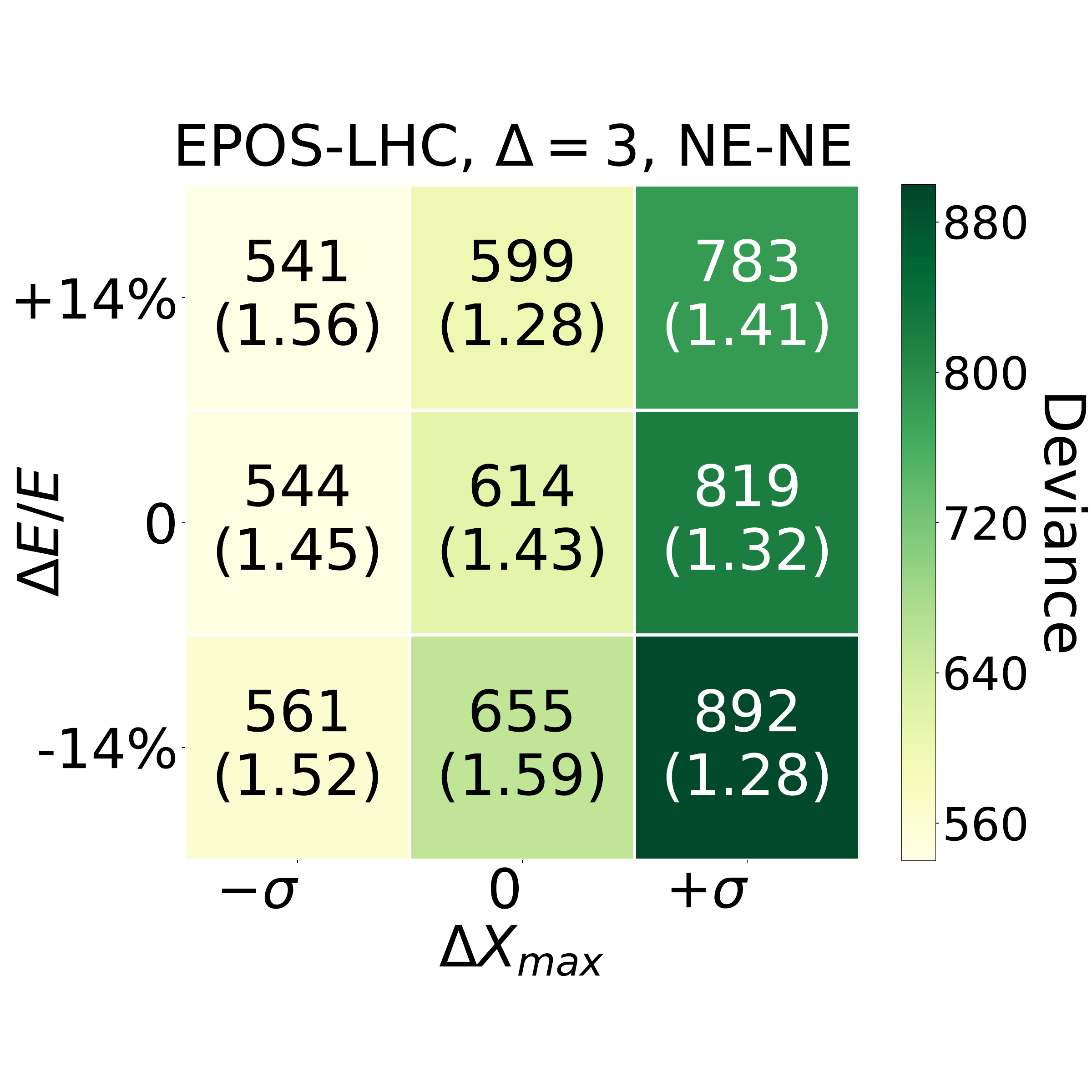}

    \includegraphics[width=0.325\textwidth]{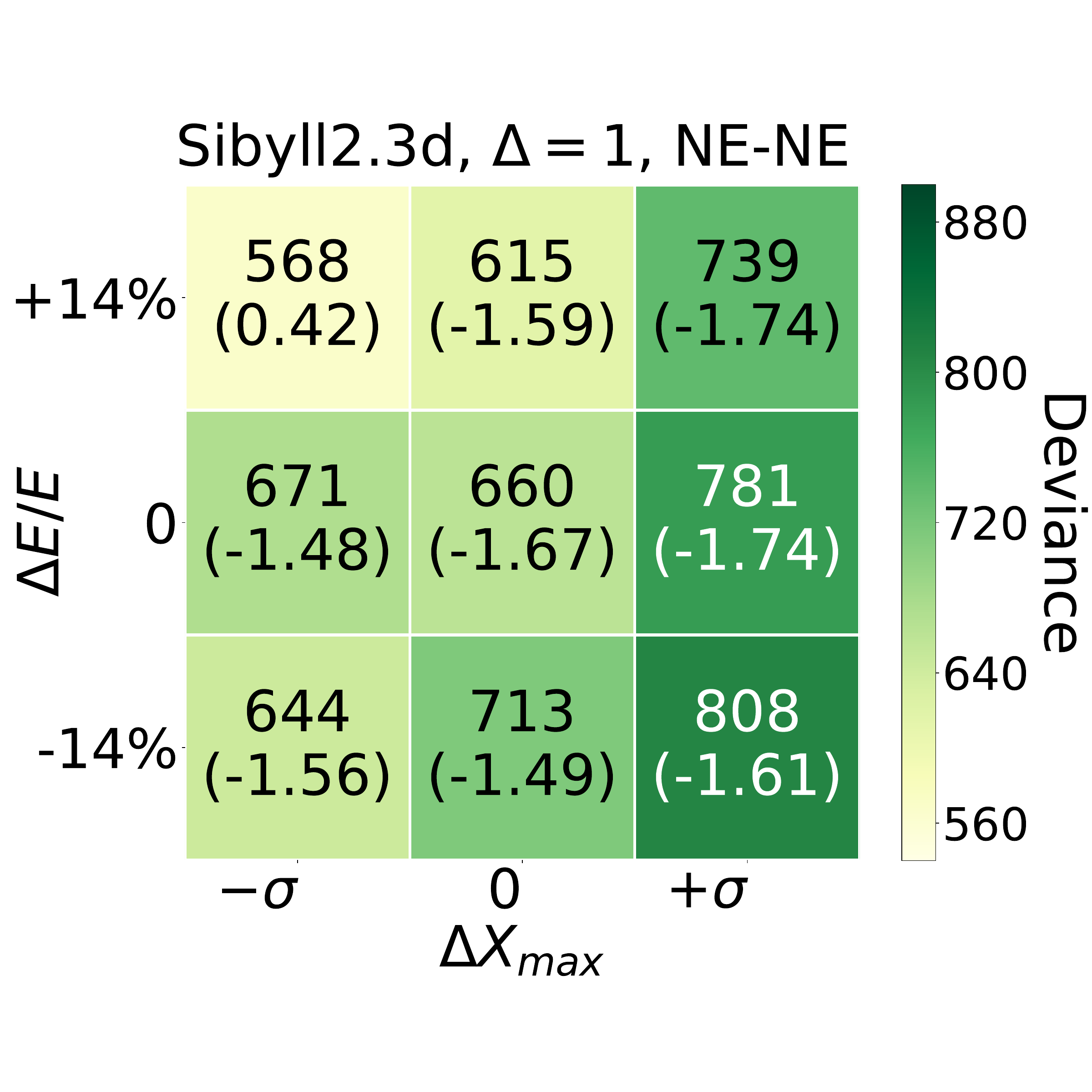}
    \includegraphics[width=0.325\textwidth]{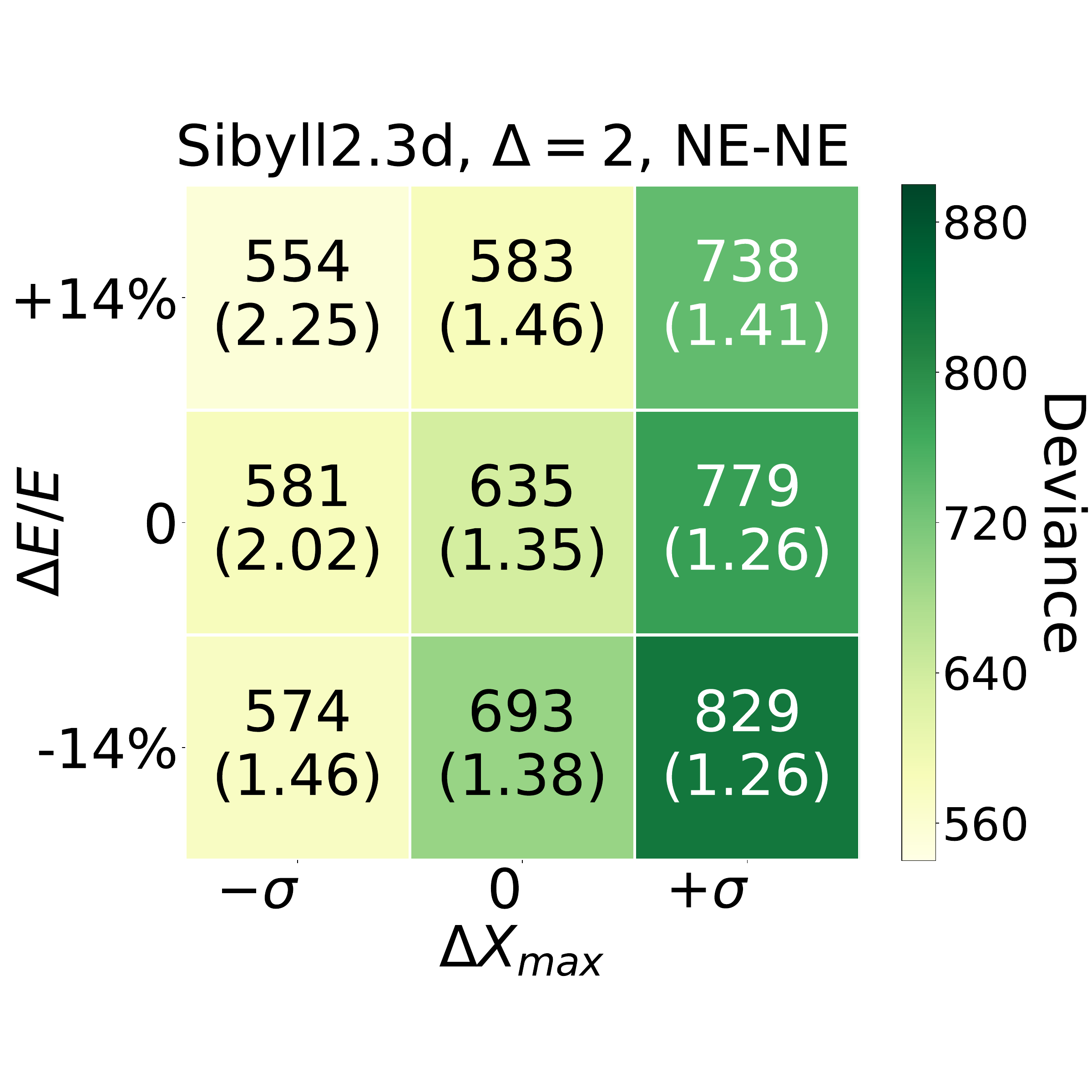}
    \includegraphics[width=0.325\textwidth]{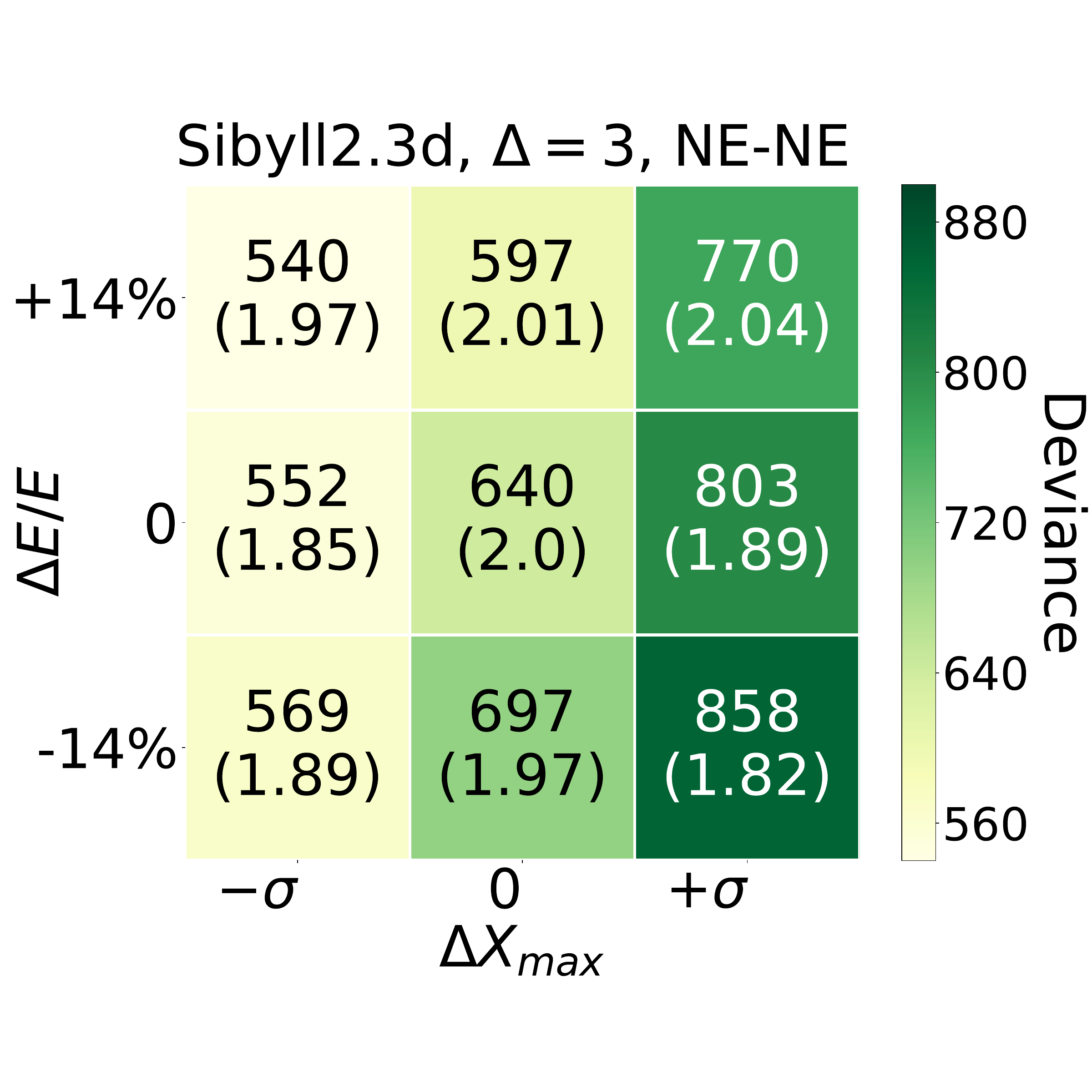}
    
    \caption{Deviance and $\gamma_{\rm H}$ (in parenthesis) resulting for shifts of $\pm \sigma_{\rm sys}$ in the energy and $X_{\rm max}$ scales for the different scenarios. }
    \label{fig:heatmaps}
\end{figure}

We explore in this section the effect that the experimental systematic uncertainties on the energy scale and $X_{\rm max}$ calibration have  on the fit results. For the energy scale, an energy independent uncertainty $\Delta E/E = \pm 14\%$ is considered in the whole energy range analysed \cite{spectrum}. The systematic uncertainties on the measured $X_{\rm max}$ values are asymmetric and slightly energy-dependent, ranging
from 6 to 9 ${\rm g\,cm^{-2}}$ \cite{comp2014}. 
We quantify the effects of these uncertainties by shifting the measured energies and the inferred $X_{\rm max}$ values by one systematic standard deviation up and down, and performing the fit again for the nine possible combinations of shifted and unshifted data sets.
Figure~\ref{fig:heatmaps} displays the resulting  total deviance and the HE population spectral index obtained for the different cutoffs and hadronic models considered. In  general the deviance improves for a positive shift in energy and/or a negative shift in $X_{\rm max}$. On the other hand, a positive shift in $X_{\rm max}$ leads to a  deviance larger by more than 100 units, specially worsening the fit to the composition data. For completeness we include in Appendix C the results for the fits performed under similar shifts of energy scale and $X_{\rm max}$ in the absence of EGMF, where similar trends are observed, although generally the deviances are larger.

\begin{table}[ht]
\centering

{\small
\scalebox{0.85}{
\begin{tabular}[H]{ @{}c c | c | c c c c | c c c c c | c | c@{}}
\multicolumn{14}{c}{Sibyll 2.3d, $\Delta=3$, NE-NE}\\
\hline
\hline
$\Delta X_{\rm max}$ &  $\Delta E /E$& & $\gamma$ & $R_{\rm cut}$ & $X_{\rm s}$ & $R_{\rm crit}$ & $f_{\rm H}$ & $f_{\rm He}$ & $f_{\rm N}$ & $f_{\rm Si}$ & $f_{\rm Fe}$ &$L_{44}$ & $D$ \\ 
& & & & [$\rm{EeV}$] & & [$\rm{EeV}$] &[\%]&[\%]&[\%]&[\%]&[\%]& &$N=353$\\ 
\hline
\multirow{6}{*}{$-\sigma_{\rm sys}$} & \multirow{2}{*}{$-14\%$} & LE & 3.53 & $>26$ & --- & --- & 25.7 &12.7 & 61.6 & \phantom{+}0\phantom{.00} & \phantom{+}0\phantom{.00} & 10.0 & \multirow{2}{*}{569 }\\
&& HE & 1.89 & \phantom{00}7.11 & 3.34 & 1.72 & \phantom{+}0\phantom{.00} & \phantom{+}0\phantom{.00} & 83.9 & 11.8 & \phantom{+}4.3 & \phantom{+}8.7 \\  \cline{3-14}

& \multirow{2}{*}{$0$}& LE& 3.51 & $2.8$ & ---& --- & 26.6 & \phantom{+}4.0 & 59.6 & \phantom{+}9.8 & \phantom{+}0\phantom{.00} & 11.7 & \multirow{2}{*}{552} \\
&& HE & 1.85 & \phantom{00}7.88 & $>3.8$ & 1.30 & \phantom{+}0\phantom{.00} & \phantom{+}0\phantom{.00} & 70.5 & 24.8 & \phantom{+}4.7 & \phantom{+}9.5 \\  \cline{3-14}

&  \multirow{2}{*}{$+14\%$} & LE & 3.49 & $>34$ & --- & --- & 24.1 & \phantom{+}8.3 & 40.4 & 27.2 & \phantom{+}0\phantom{.00} & 13.2 & \multirow{2}{*}{540}\\
&& HE & 1.97 & \phantom{00}8.75 & $>3.2$ & 1.39 &\phantom{+}0\phantom{.00} & \phantom{+}0\phantom{.00} & 59.7 & 33.1 & \phantom{+}7.2 & 13.7\\ \cline{3-14}

\hline
\hline
\multirow{6}{*}{$0$} & \multirow{2}{*}{$-14\%$} & LE & 3.66 & $>26$ & --- & --- & 18.1& 60.0 & 21.9 & \phantom{+}0\phantom{.00} & \phantom{+}0\phantom{.00} & \phantom{+}9.5 & \multirow{2}{*}{696 }\\
&& HE & 1.97 & \phantom{00}6.69 & $>3.5$ & 2.12 &\phantom{+}0\phantom{.00} & 14.2 & 73.7 & \phantom{+}8.5 & \phantom{+}3.6 & 15.1\\  \cline{3-14}

& \multirow{2}{*}{$0$}& LE & 3.62 & $>30$ & --- & --- & 21.9 & 38.1 & 39.0 & \phantom{+}0\phantom{.00} & \phantom{+}0\phantom{.00}& 11.20 & \multirow{2}{*}{640} \\
&& HE & 2.00 & \phantom{00}7.50 & 2.6 & 3.77 & \phantom{+}0\phantom{.00} & \phantom{+}4.89 & 75.4 & 15.3 & \phantom{+}4.4 & 18.4\\ \cline{3-14}

&  \multirow{2}{*}{$+14\%$} & LE & 3.60\phantom{0} & $>63$ & --- & --- & 27.4 & 16.8 & 55.8 & \phantom{+}0\phantom{.00} & \phantom{+}0\phantom{.00} & 13.0 &\multirow{2}{*}{597}\\
&& HE & 2.01 & \phantom{00}8.17 & 2.1\phantom{0} & 5.10 & \phantom{+}0.9 & \phantom{+}0\phantom{.00} & 69.6 & 24.2 & \phantom{+}5.3 & 22.6\\

\hline
\hline
\multirow{6}{*}{$+\sigma_{\rm sys}$}& \multirow{2}{*}{$-14\%$} & LE & 3.73 & $>33$ & --- & --- & 24.9 & 75.1 & \phantom{+}0\phantom{.00} & \phantom{+}0\phantom{.00} & \phantom{+}0\phantom{.00} & \phantom{+}9.5 &\multirow{2}{*}{858}\\
&& HE & 1.82 & \phantom{00}6.92 & $>3.8$ & 2.73 & \phantom{+}0\phantom{.00} & 17.7 & 76.9 & \phantom{+}2.4 & \phantom{+}3.0 & 15.2\\ \cline{3-14}

&  \multirow{2}{*}{$0$} & LE & 3.72 & $>39$  & --- & --- & 18.7 & 70.8 & 10.5 & \phantom{+}0\phantom{.00} & \phantom{+}0\phantom{.00} & 10.9 &\multirow{2}{*}{803}\\
&& HE & 1.89 & \phantom{00}7.40 & $>2.7$ & 2.77 &\phantom{+}0\phantom{.00} & 10.7 & 76.0 & \phantom{+}9.2 & \phantom{+}4.1 & 21.1\\  \cline{3-14}

&  \multirow{2}{*}{$+14\%$} & LE & 3.76 & $>39$ & --- & --- & 20.7 & 52.7 & 26.6 & \phantom{+}0\phantom{.00} & \phantom{+}0\phantom{.00} & 12.4 &\multirow{2}{*}{770 }\\ 
&& HE & 2.04 & \phantom{00}7.80 & $>2.3$ & 2.94 & \phantom{+}0\phantom{.00} & \phantom{+}5.6 & 74.6 & 14.0 & \phantom{+}5.8 & 33.9\\

\end{tabular}}}

\caption{Effects of systematic uncertainties: results for shifts of $\pm \sigma_{\rm sys}$ in the energy and $X_{\rm max}$ scales.}
\label{table_shift}
\end{table}

Since Sibyll~2.3d with $\Delta=3$ leads to the smaller deviance when performing a positive  1 $\sigma$ shift in the energy and a negative  1 $\sigma$ shift in $X_{\rm max}$, with a $\gamma_{\rm H} \simeq 2$, we display in Table~\ref{table_shift} a more in-depth exploration of the effect of the systematic shifts on the fit parameters for this scenario.
As expected, the rigidity cutoff of the high-energy component moves to larger values for a positive energy shift (and to smaller values for a negative shift), while $\gamma_{\rm H} \simeq 2$ holds in all cases. The low energy spectral parameters remain practically unchanged for most of the shifts.
On the other side, a positive shift in $X_{\rm max}$ leads to a heavier composition at the sources. 
The deviance improves by about 100 units for a positive shift in energy and a negative shift in the $X_{\rm max}$ values. For these shifts  the main thing to highlight is that one obtains $X_{\rm s} R_{\rm crit} \approx 5 \,\rm{EeV}$,  allowing hence for smaller magnetic field values and/or intersource distances than in the reference case.

\begin{figure}[ht]
    \centering
    
    \includegraphics[width=0.48\textwidth]{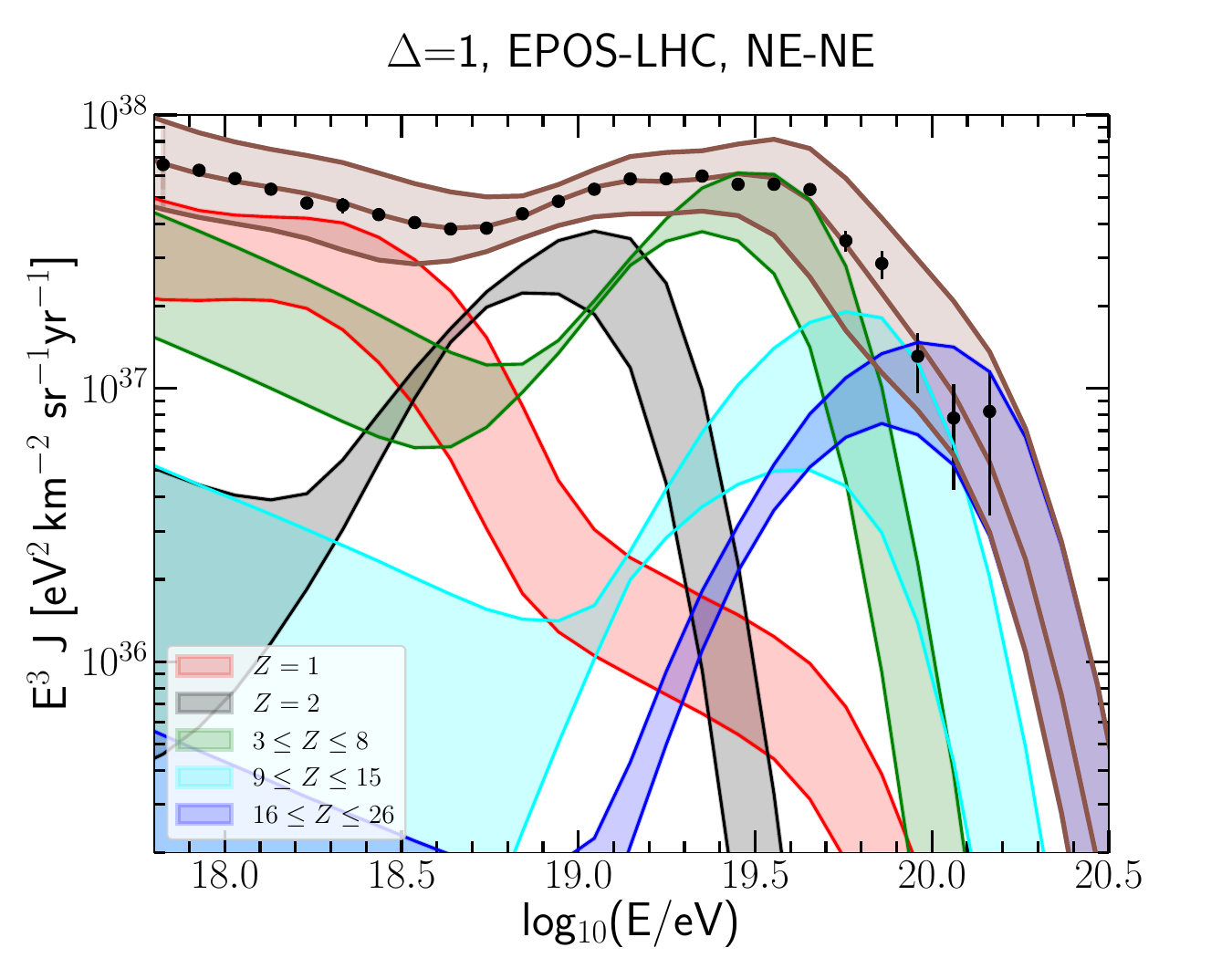}
    \includegraphics[width=0.48\textwidth]{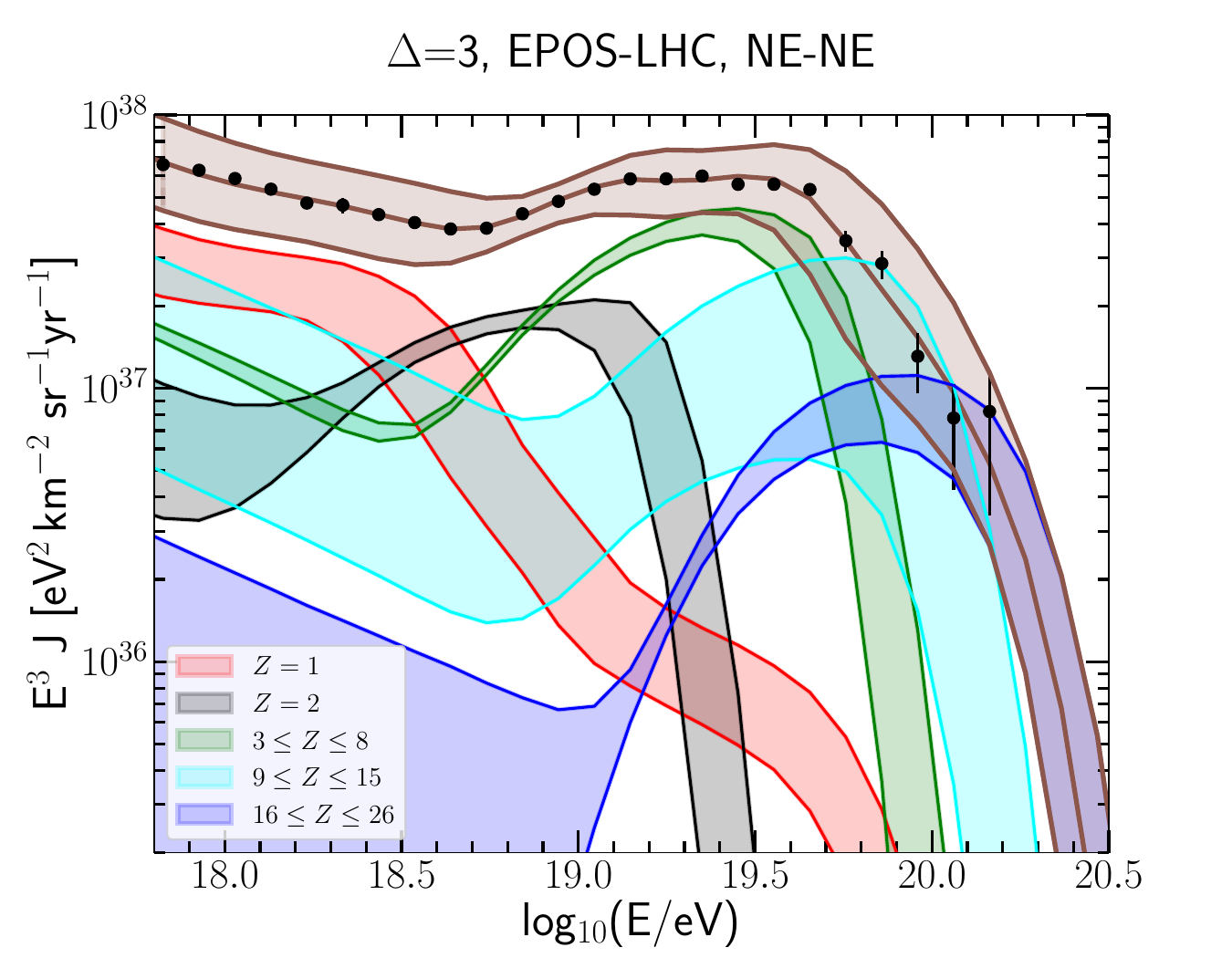}

    \includegraphics[width=0.48\textwidth]{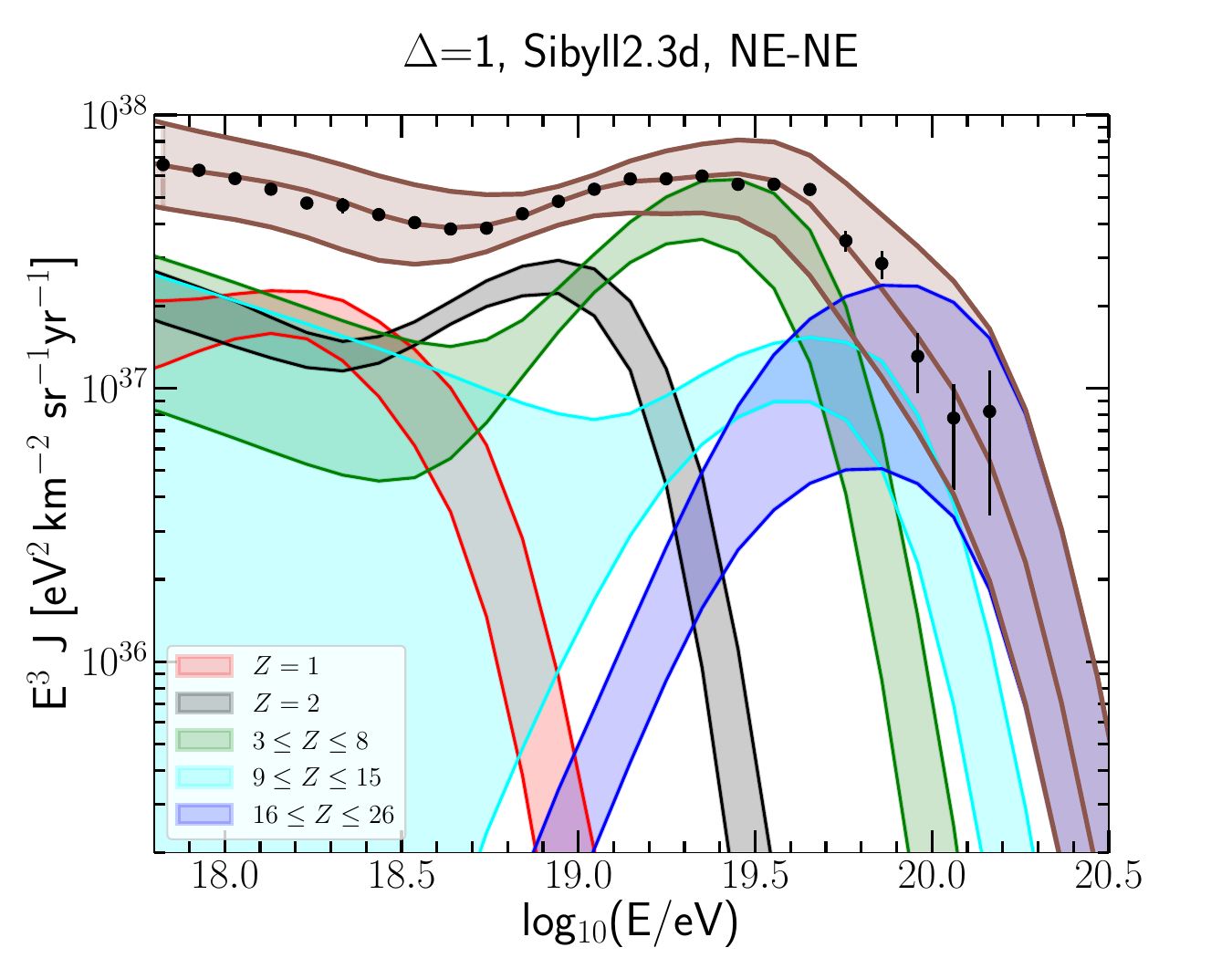}
    \includegraphics[width=0.48\textwidth]{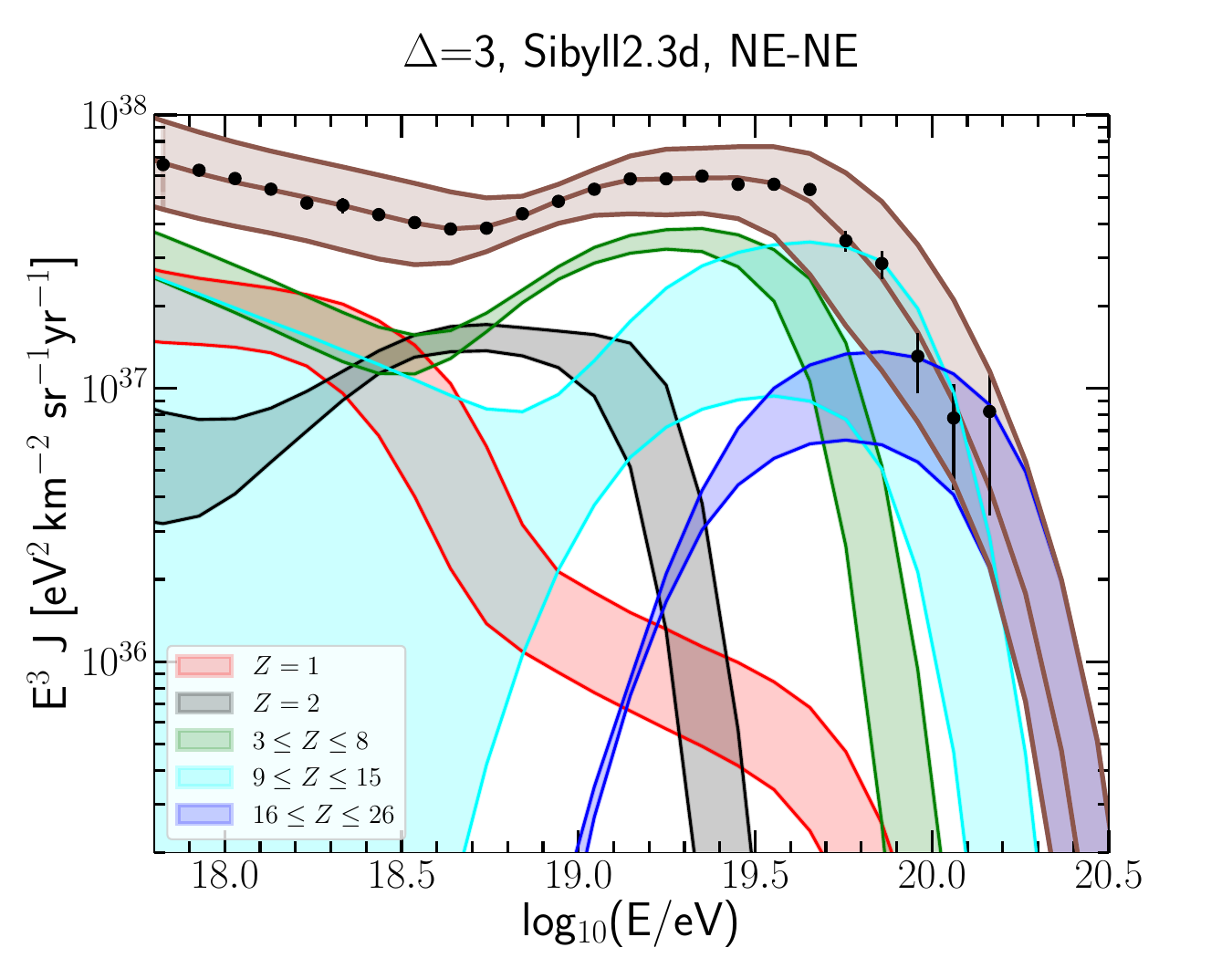}
    
    \caption{Systematic uncertainties for $\Delta E/E = 0, \pm14\%$ and the best fitting $\Delta X_{\rm max}$ for each case.}
    \label{fig:bands}
\end{figure}

Figure~\ref{fig:bands} provides a glimpse of the impact of the systematic shifts on the individual mass group fluxes, showing the typical ranges of their variations under different shifts. One of the qualitative changes that can result regards the instep feature,  that as was mentioned on section~\ref{sec_result_MHE} generally   arose from a bump in the He contribution. However,  for Sibyll~2.3d with $\Delta=3$  it actually arises mostly from the shape of the N flux, specially when a negative shift on $X_{\rm max}$ is performed. It is  seen that there are large systematic uncertainties on the LE flux of both Si and Fe, although those contributions are in general subdominant. 

\section{Conclusions}

 The combined fit to the spectrum and composition data above 0.6 EeV measured by the Pierre Auger Observatory in scenarios with two populations of sources and no extragalactic magnetic fields, which were discussed at length in (2), requires  that  the source spectrum  of the  high-energy population should be very hard, with $\gamma_{\rm H}<1$  (see Table 1). Moreover, the fit depends on the shape of the cutoff, with the spectrum being softer and the  deviance becoming smaller for milder cutoff shapes, and one obtains an extremely hard spectrum with $\gamma_{\rm H}<-1.5$ for $\Delta=1$, corresponding to the best fit.
 
On the other hand, magnetic fields are ubiquitous in the Universe, and although their strength is poorly known, they are expected to be enhanced in large scale structures, in particular in the Local Supercluster. Given  the discreteness of the source distribution, a magnetic horizon effect can significantly modify the observed spectrum because the CRs may not be able to reach us at low rigidities, making the observed spectrum harder than the source one. 
We hence explored in this paper the impact on the combined fit results when taking into account  the magnetic horizon effect. We included this effect only for the high-energy component, given
its assumed lower source density, considering that the associated effects in the denser low-energy
component happen at energies below those considered in the analysis. In particular, if the intersource
separation of the low-energy component were an order of magnitude smaller than that of the high-
energy one, the magnetic suppression effects on it would be very mild above 1 EeV. Let us note
that taking into account that the inferred emissivity per unit volume and time for the low and
high energy populations are of comparable magnitude, one expects in that case that the typical
individual luminosities of the much more abundant LE sources be fainter by at least three orders
of magnitude with respect to those of the HE sources.

\begin{figure}[t]
    \centering
 
    \includegraphics[width=0.6\textwidth]{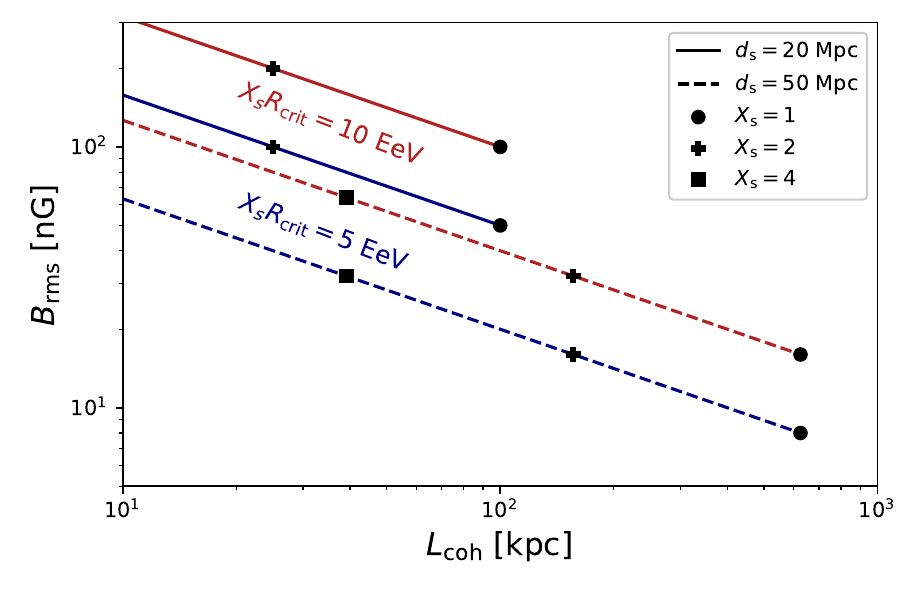}
    \caption{Magnetic field amplitude vs. coherence length required for values of the product $X_{\rm s} R_{\rm crit}$   equal to 10 EeV (red) and 5 EeV (blue). Two different values of the intersource distance are considered, $d_{\rm s} = 20$ Mpc in solid lines and 50~Mpc in dashed lines (corresponding to source densities of $\sim 10^{-4}$ and $10^{-5} {\rm Mpc}^{-3}$ respectively). Different markers indicate the values corresponding to $X_{\rm s} = 1,2$ and 4.}
    \label{fig:BvsLc}
\end{figure}

We found that,  in order for the magnetic horizon effect to play a relevant role, the normalised intersource distance should satisfy $X_{\rm s}>1$, and hence
\begin{equation}
    d_{\rm s}>20\,{\rm Mpc}\sqrt{\frac{L_{\rm coh}}{100\,{\rm kpc}}}.
\end{equation}
  Larger values of $X_{\rm s}$ require smaller critical rigidities, and the deviance is almost degenerate for $X_{\rm s} > 2$. For the scenarios where the magnetic horizon effect is responsible for the hardness of the inferred HE spectra we obtained that, allowing for the possible systematic shifts on the measurements that were discussed in the previous section,  the   approximate relation $ X_{\rm s} R_{\rm crit} \simeq 5\, {\rm to}\, 10 \ {\rm EeV} $ should hold, where  these quantities are related through 
\begin{equation}
    X_{\rm s} R_{\rm crit} \simeq 5 \ {\rm EeV} \frac{d_{\rm s}}{20\,\rm{Mpc}} \frac{B_{\rm rms}}{50\,\rm{nG}} \sqrt{ \frac{L_{\rm coh}}{100\,\rm{kpc}} }.
\end{equation}

The  values of $B_{\rm rms}$ which are required as a function of $L_{\rm coh}$ are displayed in Fig.~\ref{fig:BvsLc}, for two values of $d_{\rm s}$ (20 and 50~Mpc) and for  $X_{\rm s} R_{\rm crit} = 5$\,EeV (blue lines) and 10\,EeV (red lines). Along each line the values of $B_{\rm rms}$ depend on the associated value of $X_{\rm s}$ considered, and the values of $X_{\rm s}=1$, 2 and 4 are indicated in the plot with different symbols, and one should keep in mind that for $X_{\rm s}<1$ the magnetic horizon effect does not significantly affect the fit. 
The required  values of  the turbulent extragalactic magnetic field between the closest sources and the Earth should be strong, $B_{\rm rms} = {\cal O}(10$ to 200 nG). Although these values exceed the bounds coming from the lack of redshift dependence of rotation measurements from distant sources \cite{pshirkov16}, which apply to cosmological magnetic fields permeating all the universe, including the large scale structure voids, larger fields are inferred in the Local Supercluster \cite{vallee02} and in filamentary structures connecting clusters \cite{vernstrom21}. These values are also in the upper range of those expected to result from the amplification of primordial fields during the process of structure formation and the action of dynamo effects from the solenoidal turbulence developed in large-scale structures \cite{va17}. 
 The required magnetic fields decrease for larger coherence length values and also for larger intersource distances (i.e. for smaller source densities).

These scenarios lead to a softer HE source spectrum, specially when the source cutoff is steep ($\Delta\geq 2$), and values of $\gamma_{\rm H}\simeq 2$  are obtained for instance for Sibyll 2.3d and $\Delta =3$.
Although the origin of the observed spectral shape is qualitatively different from the one in scenarios with no magnetic fields, the overall features of the reconstructed CR composition and elemental spectra at the Earth are quite similar in the different scenarios, with the heavier nuclei dominating beyond the suppression energy, N nuclei dominating the fluxes at tens of EeV, He nuclei contributing to shape the instep feature and a large amount of secondary H contributing at few EeV energies. 

All in all, we have shown that if the source density is small and the extragalactic magnetic field is strong, the magnetic horizon effect can provide an alternative explanation of the very hard spectra of individual mass component at the Earth inferred at the highest energies.
Further studies of EGMF as well as CR anisotropy studies at the highest energies and the improved composition determination expected from the AugerPrime upgrade \cite{upgrade} should help to further constrain these scenarios.

\section*{Appendix A: Parameterization of the magnetic horizon effect}

 \begin{table}[t]
\centering
  \renewcommand{\arraystretch}{1}
 \begin{tabular}{>{\centering}m{3cm} >{\centering}m{2.1cm} >{\centering}m{3cm} >{\centering}m{2cm} >{\centering\arraybackslash}m{2cm}}

\hline\hline
NE& $a$ &  $b$ & $\alpha$ & $\beta$\\ 
\hline

Primaries & $ 0.206 + 0.026\gamma  $ & $ 0.146 + 0.004\gamma  $& $ 1.83 - 0.08\gamma$ &   \\
\cline{1-4}
Secondary  \\* \vspace{-.3\baselineskip}protons& $ 0.098 $ & $ 0.072 - 0.005\gamma  $& $ 2.02 $ & 0.129 \\
\cline{1-4}
Intermediate \\* \vspace{-.3\baselineskip}secondary nuclei & $ 0.117 $ & $ 0.092 - 0.008\gamma $& $ 2.08 $ &\\

\hline\hline
SFR& $a$ &  $b$ & $\alpha$ & $\beta$\\ 
\hline
Primaries & $ 0.135 + 0.040\gamma $ & $ 0.254 + 0.040\gamma $& $ 2.03 - 0.11\gamma $  & \\ 
\cline{1-4}
Secondary  \\* \vspace{-.3\baselineskip} protons & $ 0.117 $ & $ 0.266 - 0.029\gamma $& $ 1.99 $  &  0.29 \\ 
\cline{1-4}
Intermediate \\* \vspace{-.3\baselineskip}secondary nuclei & $ 0.103 $ & $ 0.242 - 0.040\gamma $& $ 2.01 $ & \\ 
\end{tabular}
\caption{Parameters of the fit for the magnetic suppression factor $G$, using eq.~(\ref{gfactor}),  for primary nuclei, secondary protons and intermediate mass secondary nuclei, for both the NE and SFR scenarios.}
\label{table_gparams}
\end{table}

As shown in eq.~(\ref{gfactor}), the low-energy suppression of the flux at Earth resulting from the magnetic horizon effect can be parameterised via a simple functional form that depends on the density of the sources via the parameter $X_{\rm s}$  and on the critical rigidity of the EGMF. This functional form also depends on the four parameters $a,b,\alpha$ and $\beta$, which in turn are functions of the spectral index $\gamma$, the source evolution and the kind of nuclei considered. The value of the parameters, obtained in  \cite{go21}, are reported in Table~\ref{table_gparams}.

Primaries refer to nuclei that reach Earth as part of the same mass group as that in which they were injected. Secondary protons are H nuclei that were produced via photodisintegration processes. Intermediate secondary nuclei refers to secondary nuclei, produced in photodisintegrations, belonging to a lighter mass group than the primary nuclei.

\section*{Appendix B: Impact of the shape of the cutoff}

\begin{figure}[ht]
    \centering
    \includegraphics[width=0.49\textwidth]{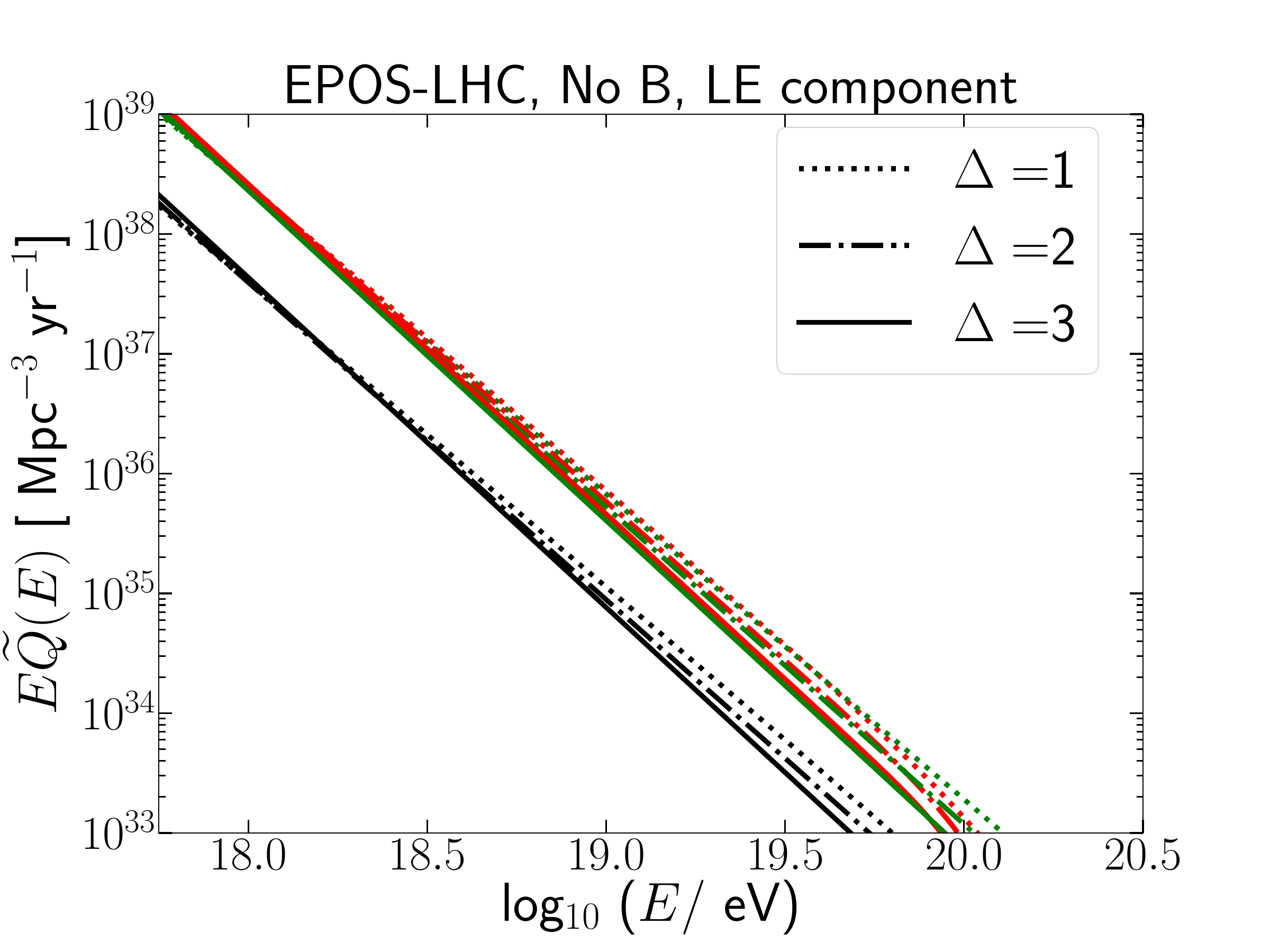}
    \includegraphics[width=0.49\textwidth]{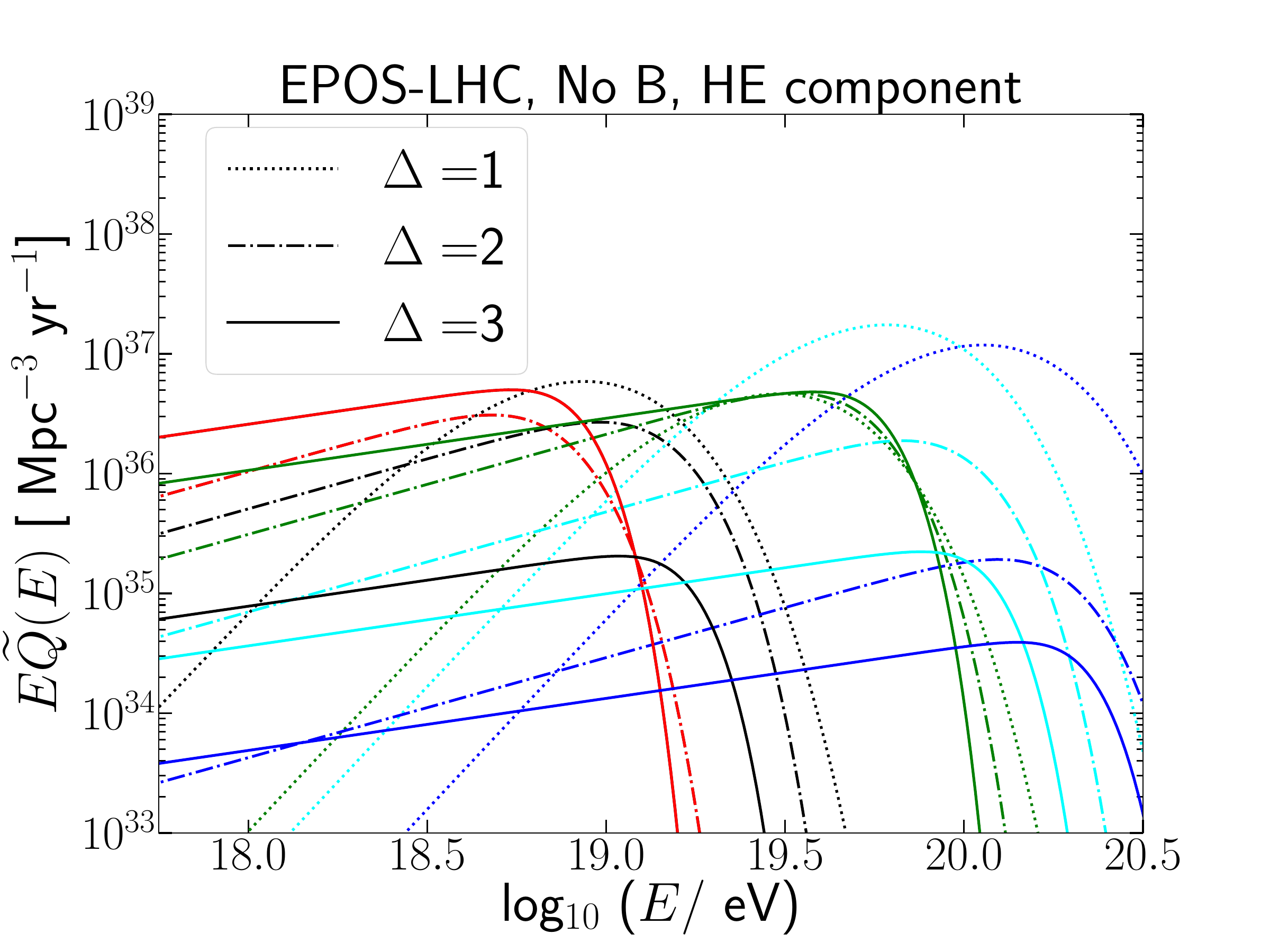}
    \caption{Spectrum of particles injected at the sources for the low-energy (left) and high-energy (right) populations, for different cutoff shapes and for the case of EPOS-LHC with no evolution and in the absence of EGMF. The H spectrum is indicated in red, the He one in gray, N  in green, Si  in cyan and Fe in blue.}
    \label{fig:injection_noB}
\end{figure}

The sharpness of the cutoff function, that is modelled with the parameter $\Delta$ in $F_{\rm cut}$, impact the results of the fit specially  for the HE spectral parameters $\gamma_{\rm H}$ and $R^{\rm H}_{\rm cut}$, even in the absence of magnetic horizon effects, as it is apparent from Table~\ref{table_0}. In Fig.~\ref{fig:injection_noB} we show the spectra of the emitted particles of the LE and HE populations for the different masses and for the three cutoff shapes considered.
Despite the large difference in the values of the parameters  for the different cutoffs, the curves look quite similar in the energy range where each mass component is dominant.

\section*{Appendix C: Effect of systematic uncertainties without magnetic horizon}

\begin{figure}[ht]
    \centering
    
    \includegraphics[width=0.325\textwidth]{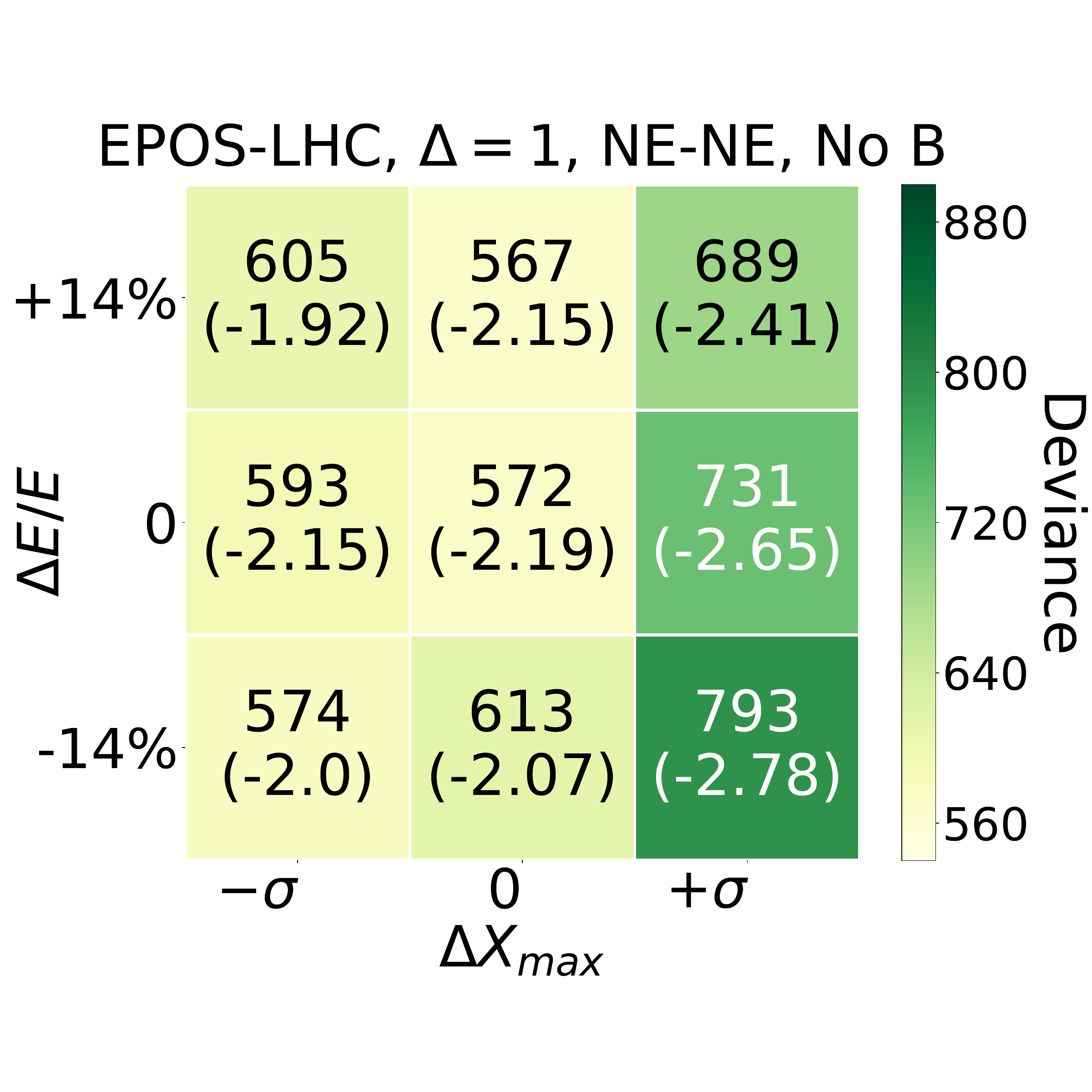}
    \includegraphics[width=0.325\textwidth]{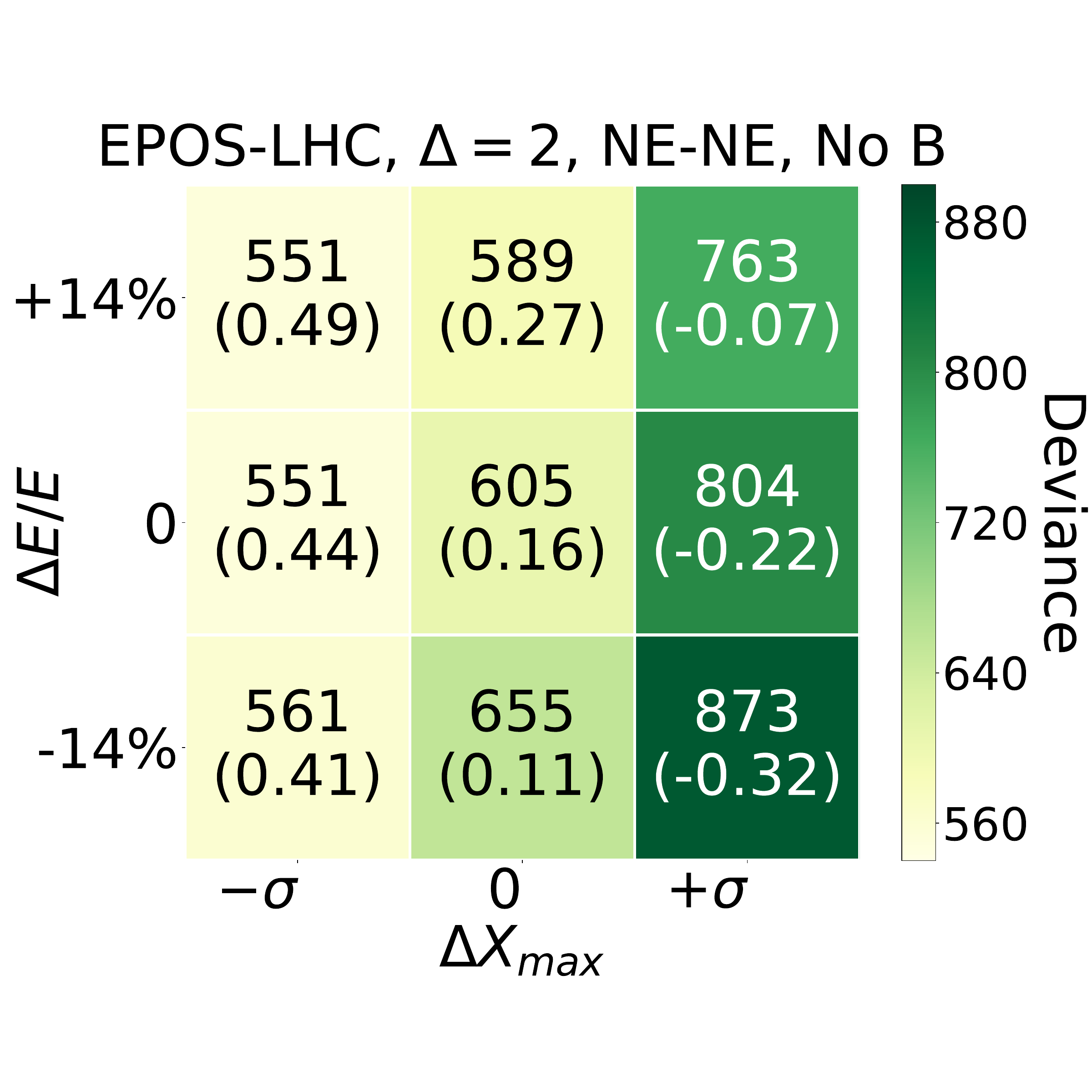}
    \includegraphics[width=0.325\textwidth]{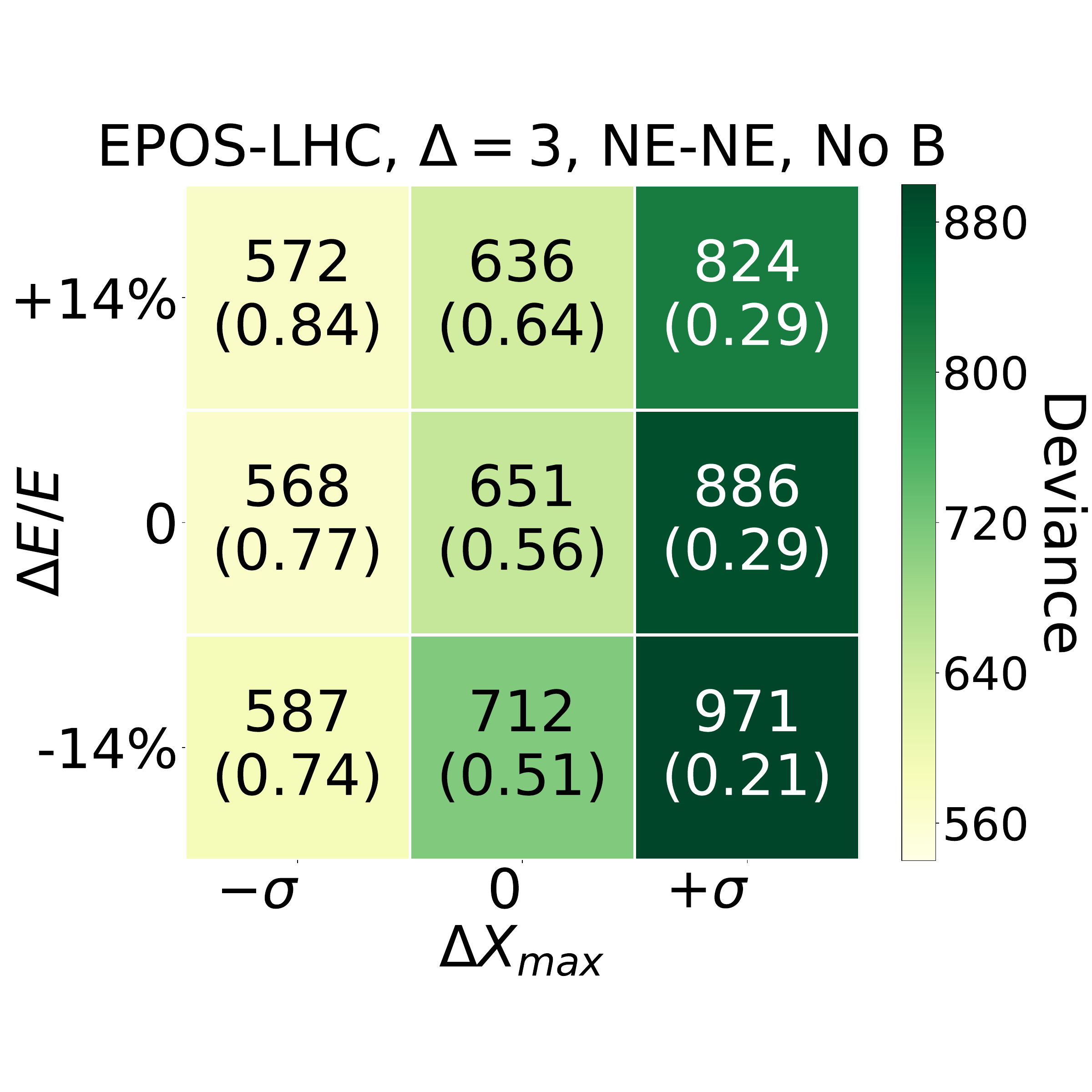}

    \includegraphics[width=0.325\textwidth]{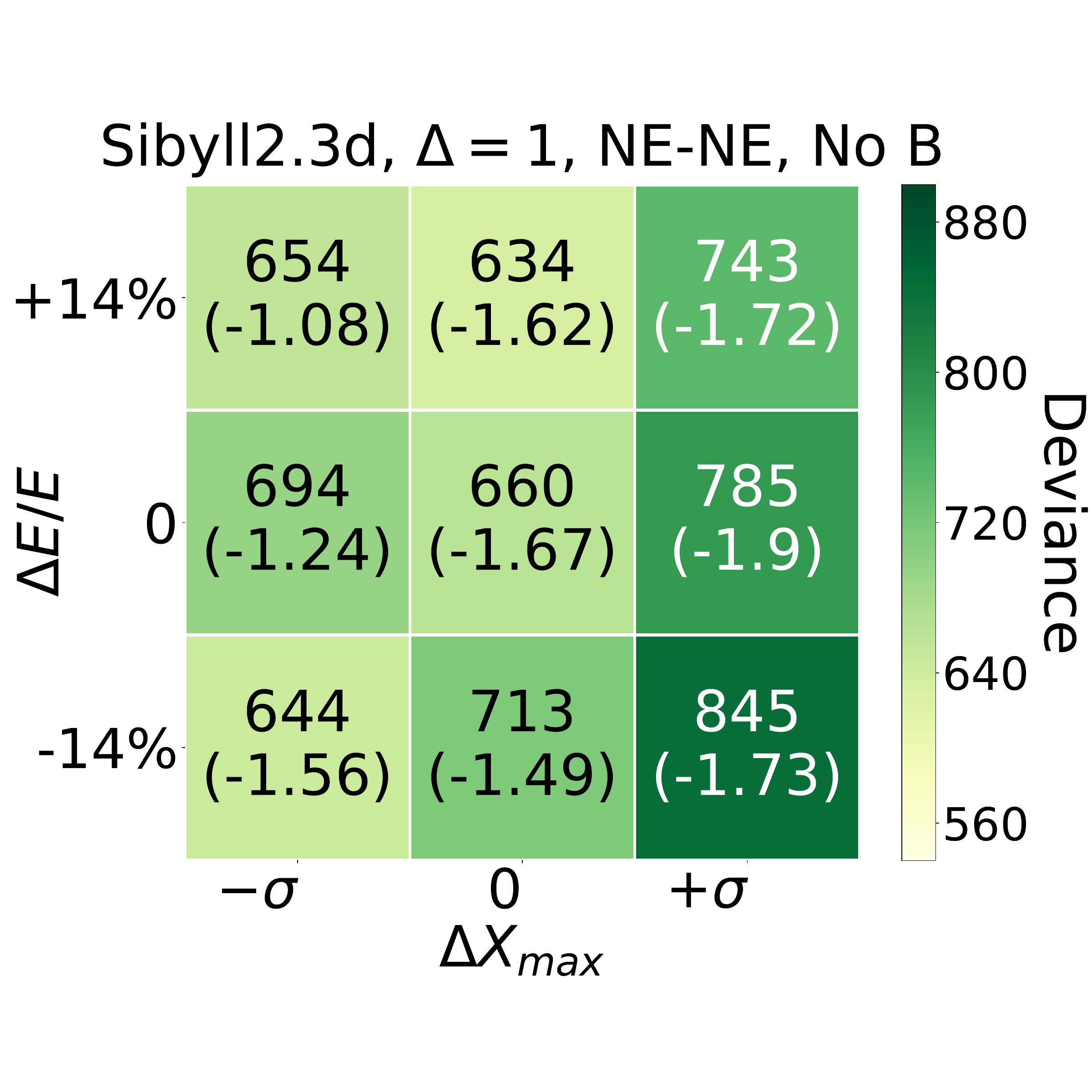}
    \includegraphics[width=0.325\textwidth]{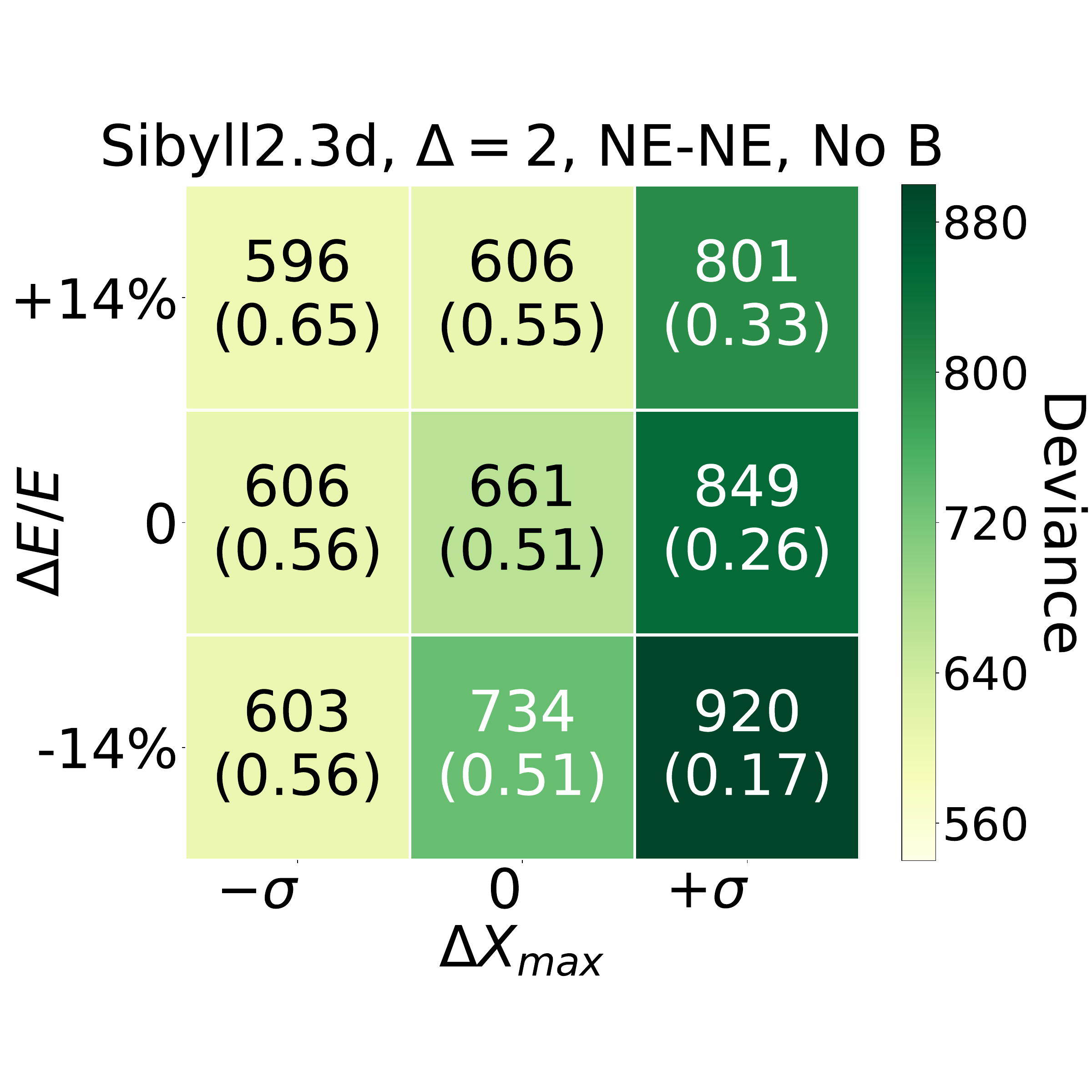}
    \includegraphics[width=0.325\textwidth]{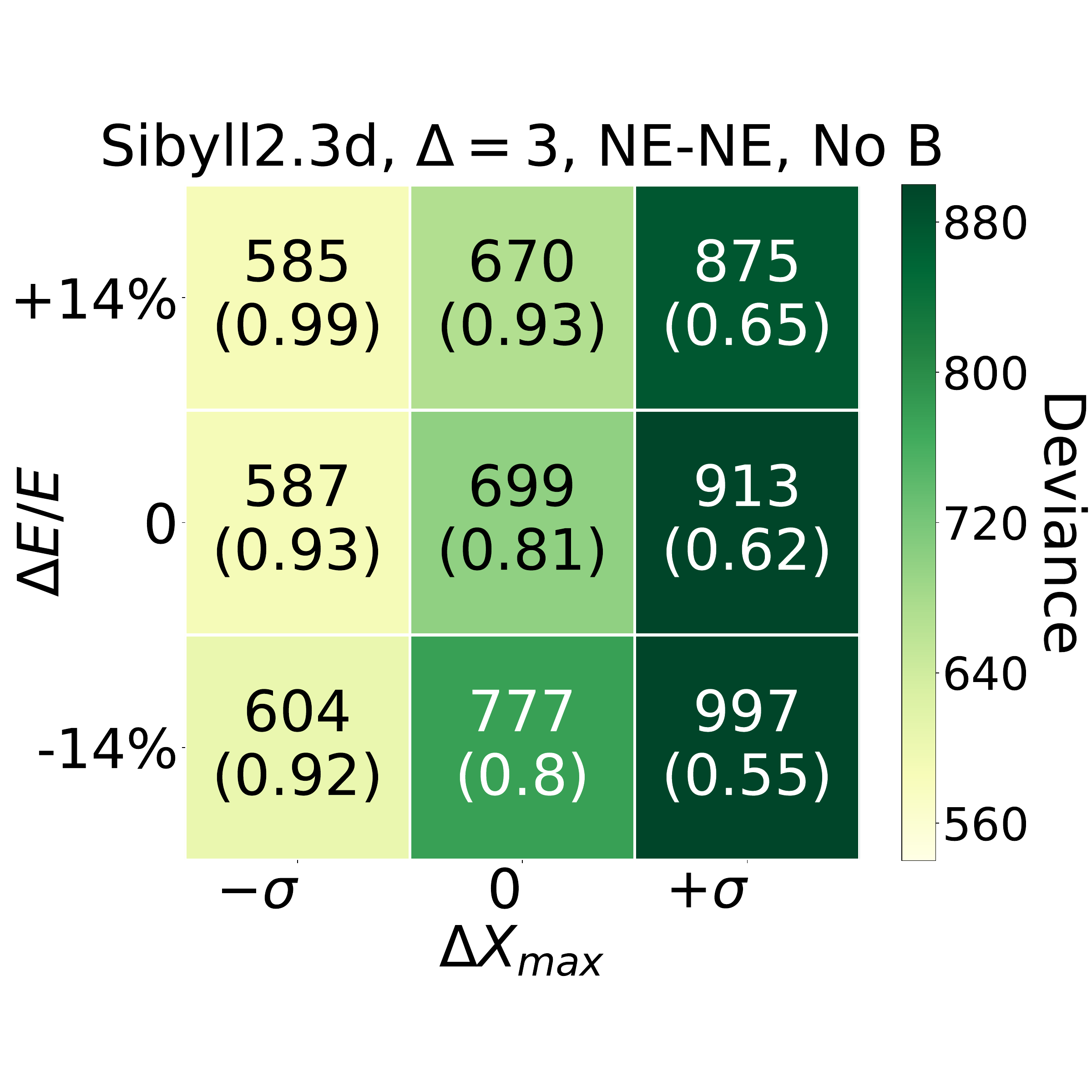}
    
    \caption{Deviance and $\gamma_{\rm H}$ (in parenthesis) resulting for shifts of $\pm \sigma_{\rm sys}$ in the energy and $X_{\rm max}$ scales for the different scenarios in the absence of EGMF. }
    \label{fig:heatmapsnoB}
\end{figure}

We report here the effect on the fits resulting from shifting the measured energies and the inferred $X_{\rm max}$ values by one systematic standard deviation up and down. Fig.~\ref{fig:heatmapsnoB} displays the deviance and the HE spectral index (in parenthesis) for the nine possible combinations of shifts, analogous to those in Fig.~\ref{fig:heatmaps} for the case with magnetic horizon effects.

\newpage
\section*{Acknowledgments}

\begin{sloppypar}
The successful installation, commissioning, and operation of the Pierre
Auger Observatory would not have been possible without the strong
commitment and effort from the technical and administrative staff in
Malarg\"ue. We are very grateful to the following agencies and
organizations for financial support:
\end{sloppypar}

\begin{sloppypar}
Argentina -- Comisi\'on Nacional de Energ\'\i{}a At\'omica; Agencia Nacional de
Promoci\'on Cient\'\i{}fica y Tecnol\'ogica (ANPCyT); Consejo Nacional de
Investigaciones Cient\'\i{}ficas y T\'ecnicas (CONICET); Gobierno de la
Provincia de Mendoza; Municipalidad de Malarg\"ue; NDM Holdings and Valle
Las Le\~nas; in gratitude for their continuing cooperation over land
access; Australia -- the Australian Research Council; Belgium -- Fonds
de la Recherche Scientifique (FNRS); Research Foundation Flanders (FWO),
Marie Curie Action of the European Union Grant No.~101107047; Brazil --
Conselho Nacional de Desenvolvimento Cient\'\i{}fico e Tecnol\'ogico (CNPq);
Financiadora de Estudos e Projetos (FINEP); Funda\c{c}\~ao de Amparo \`a
Pesquisa do Estado de Rio de Janeiro (FAPERJ); S\~ao Paulo Research
Foundation (FAPESP) Grants No.~2019/10151-2, No.~2010/07359-6 and
No.~1999/05404-3; Minist\'erio da Ci\^encia, Tecnologia, Inova\c{c}\~oes e
Comunica\c{c}\~oes (MCTIC); Czech Republic -- GACR 24-13049S, CAS LQ100102401,
MEYS LM2023032, CZ.02.1.01/0.0/0.0/16{\textunderscore}013/0001402,
CZ.02.1.01/0.0/0.0/18{\textunderscore}046/0016010 and
CZ.02.1.01/0.0/0.0/17{\textunderscore}049/0008422 and CZ.02.01.01/00/22{\textunderscore}008/0004632;
France -- Centre de Calcul IN2P3/CNRS; Centre National de la Recherche
Scientifique (CNRS); Conseil R\'egional Ile-de-France; D\'epartement
Physique Nucl\'eaire et Corpusculaire (PNC-IN2P3/CNRS); D\'epartement
Sciences de l'Univers (SDU-INSU/CNRS); Institut Lagrange de Paris (ILP)
Grant No.~LABEX ANR-10-LABX-63 within the Investissements d'Avenir
Programme Grant No.~ANR-11-IDEX-0004-02; Germany -- Bundesministerium
f\"ur Bildung und Forschung (BMBF); Deutsche Forschungsgemeinschaft (DFG);
Finanzministerium Baden-W\"urttemberg; Helmholtz Alliance for
Astroparticle Physics (HAP); Helmholtz-Gemeinschaft Deutscher
Forschungszentren (HGF); Ministerium f\"ur Kultur und Wissenschaft des
Landes Nordrhein-Westfalen; Ministerium f\"ur Wissenschaft, Forschung und
Kunst des Landes Baden-W\"urttemberg; Italy -- Istituto Nazionale di
Fisica Nucleare (INFN); Istituto Nazionale di Astrofisica (INAF);
Ministero dell'Universit\`a e della Ricerca (MUR); CETEMPS Center of
Excellence; Ministero degli Affari Esteri (MAE), ICSC Centro Nazionale
di Ricerca in High Performance Computing, Big Data and Quantum
Computing, funded by European Union NextGenerationEU, reference code
CN{\textunderscore}00000013; M\'exico -- Consejo Nacional de Ciencia y Tecnolog\'\i{}a
(CONACYT) No.~167733; Universidad Nacional Aut\'onoma de M\'exico (UNAM);
PAPIIT DGAPA-UNAM; The Netherlands -- Ministry of Education, Culture and
Science; Netherlands Organisation for Scientific Research (NWO); Dutch
national e-infrastructure with the support of SURF Cooperative; Poland
-- Ministry of Education and Science, grants No.~DIR/WK/2018/11 and
2022/WK/12; National Science Centre, grants No.~2016/22/M/ST9/00198,
2016/23/B/ST9/01635, 2020/39/B/ST9/01398, and 2022/45/B/ST9/02163;
Portugal -- Portuguese national funds and FEDER funds within Programa
Operacional Factores de Competitividade through Funda\c{c}\~ao para a Ci\^encia
e a Tecnologia (COMPETE); Romania -- Ministry of Research, Innovation
and Digitization, CNCS-UEFISCDI, contract no.~30N/2023 under Romanian
National Core Program LAPLAS VII, grant no.~PN 23 21 01 02 and project
number PN-III-P1-1.1-TE-2021-0924/TE57/2022, within PNCDI III; Slovenia
-- Slovenian Research Agency, grants P1-0031, P1-0385, I0-0033, N1-0111;
Spain -- Ministerio de Ciencia e Innovaci\'on/Agencia Estatal de
Investigaci\'on (PID2019-105544GB-I00, PID2022-140510NB-I00 and
RYC2019-027017-I), Xunta de Galicia (CIGUS Network of Research Centers,
Consolidaci\'on 2021 GRC GI-2033, ED431C-2021/22 and ED431F-2022/15),
Junta de Andaluc\'\i{}a (SOMM17/6104/UGR and P18-FR-4314), and the European
Union (Marie Sklodowska-Curie 101065027 and ERDF); USA -- Department of
Energy, Contracts No.~DE-AC02-07CH11359, No.~DE-FR02-04ER41300,
No.~DE-FG02-99ER41107 and No.~DE-SC0011689; National Science Foundation,
Grant No.~0450696; The Grainger Foundation; Marie Curie-IRSES/EPLANET;
European Particle Physics Latin American Network; and UNESCO.
\end{sloppypar}


\begin{thebibliography}{00}

\bibitem{aa17} 
A. Aab et al. (Pierre Auger Collaboration), \href{https://iopscience.iop.org/article/10.1088/1475-7516/2017/04/038}{
JCAP 04 (2017) 038}

\bibitem{xcf} 
A. Abdul Halim et al. (Pierre Auger Collaboration), \href{https://iopscience.iop.org/article/10.1088/1475-7516/2023/05/024}{JCAP 05 (2023) 024}

\bibitem{Aloisio:2013hya}
R.~Aloisio, V.~Berezinsky and P.~Blasi,
\href{https://iopscience.iop.org/article/10.1088/1475-7516/2014/10/020}{JCAP 10 (2014) 020}

\bibitem{gl15}
N.~Globus, D.~Allard and E.~Parizot,
\href{https://journals.aps.org/prd/abstract/10.1103/PhysRevD.92.021302}{Phys. Rev. D 92 (2015)  021302}

\bibitem{AlvesBatista:2018zui}
R.~Alves Batista, R.M.~de Almeida, B.~Lago and K.~Kotera,
\href{https://iopscience.iop.org/article/10.1088/1475-7516/2019/01/002}{JCAP \textbf{01} (2019) 002}

\bibitem{Heinze:2019jou}
J.~Heinze, A.~Fedynitch, D.~Boncioli and W.~Winter,
\href{https://iopscience.iop.org/article/10.3847/1538-4357/ab05ce}{Astrophys. J. \textbf{873} (2019)  88}

\bibitem{ayus19} A.\ Yushkov (for the Pierre Auger Collaboration),  \href{https://pos.sissa.it/358/482/pdf}{PoS(ICRC2019)482}

\bibitem{comp2014}
A. Aab et al. (Pierre Auger Collaboration), \href{https://journals.aps.org/prd/abstract/10.1103/PhysRevD.90.122005}{Phys. Rev. D 90 (2014) 122005}

\bibitem{adcfit} 
A. Abdul Halim et al. (Pierre Auger Collaboration), \href{https://iopscience.iop.org/article/10.1088/1475-7516/2024/01/022}{JCAP 01 (2024) 022}

\bibitem{Globus:2014fka}
N.~Globus, D.~Allard, R.~Mochkovitch and E.~Parizot,
 \href{https://academic.oup.com/mnras/article/451/1/751/1354047?login=false}{Mon. Not. Roy. Astron. Soc. 451 (2015)  751}


\bibitem{un15} M. Unger, G.R. Farrar and L. Anchordoqui, 
\href{https://journals.aps.org/prd/abstract/10.1103/PhysRevD.92.123001}{Phys.\ Rev.\ D\ {92} (2015) 123001}

%
\bibitem{bi18} D. Biehl, D. Boncioli, A. Fedynitch and W. Winter,
\href{https://www.aanda.org/articles/aa/abs/2018/03/aa31337-17/aa31337-17.html}{Astron. and Astrophys. 611 (2018) A101}



\bibitem{Supanitsky:2018jje}
A.D.~Supanitsky, A.~Cobos and A.~Etchegoyen,
\href{https://journals.aps.org/prd/abstract/10.1103/PhysRevD.98.103016}{Phys. Rev. D 98 (2018) 103016}

\bibitem{Muzio:2019leu}
M.S.~Muzio, M.~Unger and G.R.~Farrar,
\href{https://journals.aps.org/prd/abstract/10.1103/PhysRevD.100.103008}{Phys. Rev. D 100 (2019)  103008}

\bibitem{he20}
J.~Heinze, D.~Biehl, A.~Fedynitch, D.~Boncioli, A.~Rudolph and W.~Winter,
\href{https://doi.org/10.1093/mnras/staa2751}{
{Mon.\ Not.\ Roy.\ Astron.\ Soc.} 498 (2020) 5990}

\bibitem{co23}
A.~Condorelli, D.~Boncioli, E.~Peretti and S.~Petrera,
\href{https://journals.aps.org/prd/abstract/10.1103/PhysRevD.107.083009}{Phys. Rev. D 107 (2023)  083009}


\bibitem{ka06} M.\ Kachelriess and D.V.\ Semikoz,   \href{https://www.sciencedirect.com/science/article/abs/pii/S0370269306000670}{Phys.\ Lett.\ B 634 (2006) 143}

\bibitem{luce22}
Q. Luce, S. Marafico, J. Biteau, A. Condorelli and O. Deligny, \href{https://iopscience.iop.org/article/10.3847/1538-4357/ac81cc}{Astrophys.\ J.\ {936} (2022) {62}}

\bibitem{al04}
 R.\ Aloisio and V.\ Berezinsky, \href{https://iopscience.iop.org/article/10.1086/421869/meta}{Astrophys.\ J.\ {612} (2004) {900}}

\bibitem{al11}
 R.\ Aloisio, V.\ Berezinsky and A. Gazizov, \href{https://www.sciencedirect.com/science/article/pii/S0927650510002434?via%3Dihub}{Astropart. Phys. 34 (2011) 620}


\bibitem{mo13} 
S. Mollerach and E. Roulet,  \href{https://iopscience.iop.org/article/10.1088/1475-7516/2013/10/013}{JCAP 10 (2013) 013}

\bibitem{wit17}
D. Wittkowski (for the Pierre Auger Collaboration), \href{https://pos.sissa.it/301/563/pdf}{PoS(ICRC2017)563}

\bibitem{mo20} 
S.\ Mollerach and E.\ Roulet, \href{https://journals.aps.org/prd/abstract/10.1103/PhysRevD.101.103024}{Phys.\ Rev.\ D  101 (2020) {103024}}

\bibitem{go23}
J. Gonz\'alez (for the Pierre Auger Collaboration), \href{https://pos.sissa.it/444/288/pdf}{PoS(ICRC2023)288}

\bibitem{foteini}
D. Ehlert, F. Oikonomou, and  M. Unger, \href{https://journals.aps.org/prd/abstract/10.1103/PhysRevD.107.103045}{Phys. Rev. D 107 (2023) 103045}

\bibitem{ho06} 
A.M. Hopkins and J.F. Beacom,  \href{https://iopscience.iop.org/article/10.1086/506610/meta}{Astrophys. J. 651 (2006) 142}

\bibitem{al17} 
R. Aloisio et al.,  \href{https://iopscience.iop.org/article/10.1088/1475-7516/2017/11/009}{JCAP 11 (2017) 009}

\bibitem{talys} 
A.J. Koning, S. Hilaire and M.C. Duijvestijn, TALYS: Comprehensive Nuclear Reaction Modeling, in
International Conference on Nuclear Data for Science and Technology (R.C. Haight, M.B. Chadwick, T. Kawano and P. Talou, eds.), \href{https://pubs.aip.org/aip/acp/article-abstract/769/1/1154/946255/TALYS-Comprehensive-Nuclear-Reaction-Modeling?redirectedFrom=fulltext}{vol. 769 of American Institute of Physics Conference Series (2005), pp. 1154}

\bibitem{gi12} R. Gilmore, R. Somerville, J. Primack and A. Dom\'inguez,  \href{https://academic.oup.com/mnras/article/422/4/3189/1050758?login=false}{Mon. Not. Roy. Astron. Soc. 422 (2012) 3189}

\bibitem{epos} T. Pierog et al., \href{https://journals.aps.org/prc/abstract/10.1103/PhysRevC.92.034906}{Phys. Rev. C 92  (2015) 034906 }

\bibitem{sibyll} 
F. Riehn, R. Engel, A. Fedynitch, T. K. Gaisser and T. Stanev,  \href{https://journals.aps.org/prd/abstract/10.1103/PhysRevD.102.063002}{Phys. Rev. D 102 (2020) 063002}

\bibitem{lemoine} M. Lemoine,  \href{https://journals.aps.org/prd/abstract/10.1103/PhysRevD.71.083007}{Phys. Rev. D 71 (2005) 083007}

\bibitem{be07} V. Berezinsky and A.Z. Gazizov, 
\href{https://iopscience.iop.org/article/10.1086/520498/meta}{Astrophys. J. 669 (2007) 684}

\bibitem{bere08} V. Berezinsky, \href{https://www.sciencedirect.com/science/article/abs/pii/S0273117707001822}{Advances in Space Research 41 (2008) 2071}

\bibitem{mo19} S.\ Mollerach and E.\ Roulet, \href{https://journals.aps.org/prd/abstract/10.1103/PhysRevD.99.103010}{Phys.\ Rev.\ D  99 (2019) {103010}}

\bibitem{ei23}    B. Eichmann and M. Kachelriess,
\href{https://iopscience.iop.org/article/10.1088/1475-7516/2023/02/053}{JCAP 02 (2023) 053}


   
\bibitem{haverkorn15}
M. Haverkorn,  \href{https://link.springer.com/chapter/10.1007/978-3-662-44625-6_17}{Astrophysics and Space Science Library (2015) 407}

\bibitem{fe12} 
L. Feretti et al., \href{https://link.springer.com/article/10.1007/s00159-012-0054-z}{The Astron. and Astrophys. Rev. 20 (2012) 54}


\bibitem{va11} 
J.P. Vall\'ee, \href{https://www.sciencedirect.com/science/article/abs/pii/S1387647311000182}{New Astronomy Reviews 55 (2011) 91}

\bibitem{xu06}
Y. Xu, P.P. Kronberg, S. Habib and Q.W. Dufton, \href{https://iopscience.iop.org/article/10.1086/498336/pdf}{Astrophys. J. 637 (2006) 19}

\bibitem{va17} 
F. Vazza et al., \href{https://iopscience.iop.org/article/10.1088/1361-6382/aa8e60}{Class. Quantum Grav. 34 (2017) 234001}

\bibitem{du13} R. Durrer and A. Neronov, \href{https://link.springer.com/article/10.1007/s00159-013-0062-7}{The Astron. and Astrophys. Rev. 21 (2013) 62}

\bibitem{pshirkov16}
M.S. Pshirkov, P.G. Tinyakov, F.R. Urban, \href{https://journals.aps.org/prl/abstract/10.1103/PhysRevLett.116.191302}{Phys. Rev. Lett. 116 (2016) 19, 191302}

\bibitem{ry98}{D. Ryu, H. Kang and P.L. Biermann, \href{https://doi.org/10.48550/arXiv.astro-ph/9803275}{Astron. and Astrophys. 335 (1998) 19}}

\bibitem{go21} 
J. Gonz\'alez, S. Mollerach and E. Roulet, \href{https://journals.aps.org/prd/abstract/10.1103/PhysRevD.104.063005}{Phys. Rev. D 104 (2021) 063005 }

\bibitem{blasi15}
P.\ Blasi, E.\ Amato and M.\ D’Angelo, \href{https://journals.aps.org/prl/abstract/10.1103/PhysRevLett.115.121101}{Phys. Rev. Lett. 115 (2015) 121101}; A. Cermenati, R. Aloisio, P. Blasi and C. Evoli, \href{https://pos.sissa.it/444/1131/pdf}{PoS(ICRC2023)444}


\bibitem{Minuit}
F. James and M. Roos,  \href{https://www.sciencedirect.com/science/article/abs/pii/0010465575900399}{Comput. Phys. Commun. 10 (1975) 343} 


\
\bibitem{spectrum}
P.\ Abreu et al. (Pierre Auger Collaboration), \href{https://link.springer.com/article/10.1140/epjc/s10052-021-09700-w}{Eur.\ Phys.\ J.\ C {81} (2021) {966}}

\bibitem{upgrade}
A. Aab et al. (Pierre Auger Collaboration), ``The Pierre Auger Observatory Upgrade - Preliminary Design Report”, (2016) \href{https://arxiv.org/abs/1604.03637}{arXiv:1604.03637 [astro-ph]}

\bibitem{vallee02}
J.P. Vall\'ee, \href{https://web.archive.org/web/20181030073332id_/http://iopscience.iop.org/article/10.1086/342281/pdf}{The Astron. J. 124 (2002) 1322}

\bibitem{vernstrom21}
T. Vernstrom et al., \href{https://academic.oup.com/mnras/article/505/3/4178/6273648}{Mon. Not. Roy. Astron. Soc. 505 (2021) 4178}



\end{thebibliography}
\end{document}